\documentclass[dvips,12pt,a4paper]{article}
\usepackage{a4wide}
\usepackage{epsfig}
\usepackage{amsmath}
\usepackage{latexsym}

  \newlength{\absize}
\newcommand{\half}{{\textstyle\frac{1}{2}}}

\renewcommand{\Re}{\mbox{Re}}

\allowdisplaybreaks 

\catcode`@=11
\def\citer{\@ifnextchar [{\@tempswatrue\@citexr}{\@tempswafalse\@citexr[]}}
 
\def\@citexr[#1]#2{\if@filesw\immediate\write\@auxout{\string\citation{#2}}\fi
  \def\@citea{}\@cite{\@for\@citeb:=#2\do
    {\@citea\def\@citea{--\penalty\@m}\@ifundefined
       {b@\@citeb}{{\bf ?}\@warning
       {Citation `\@citeb' on page \thepage \space undefined}}%
\hbox{\csname b@\@citeb\endcsname}}}{#1}}
\catcode`@=12

\begin{document}
  \thispagestyle{empty}
  \pagestyle{empty}
  \renewcommand{\thefootnote}{\fnsymbol{footnote}}
\newpage\normalsize
    \pagestyle{plain}
    \setlength{\baselineskip}{4ex}\par
    \setcounter{footnote}{0}
    \renewcommand{\thefootnote}{\arabic{footnote}}
\newcommand{\preprint}[1]{%
  \begin{flushright}
    \setlength{\baselineskip}{3ex} #1
  \end{flushright}}
\renewcommand{\title}[1]{%
  \begin{center}
    \LARGE #1
  \end{center}\par}
\renewcommand{\author}[1]{%
  \vspace{2ex}
  {\Large
   \begin{center}
     \setlength{\baselineskip}{3ex} #1 \par
   \end{center}}}
\renewcommand{\thanks}[1]{\footnote{#1}}
\renewcommand{\abstract}[1]{%
  \vspace{2ex}
  \normalsize
  \begin{center}
    \centerline{\bf Abstract}\par
    \vspace{2ex}
    \parbox{\absize}{#1\setlength{\baselineskip}{2.5ex}\par}
  \end{center}}

\hyphenation{phenomeno-logy}
\renewcommand{\thefootnote}{\fnsymbol{footnote}}
\begin{flushright}
{\setlength{\baselineskip}{2ex}\par
{February 2003}           \\
} 
\end{flushright}
\vspace*{4mm}
\vfill
\title{Graviton-induced Bremsstrahlung}
\vfill
\author{
Erik Dvergsnes$^{a,}$\footnote{\tt erik.dvergsnes@fi.uib.no},
Per Osland$^{a,}$\footnote{\tt per.osland@fi.uib.no},
Nurcan \"Ozt\"urk$^{a,b,}$\footnote{\tt nurcan@uta.edu}}
\begin{center}
$^{a}$Department of Physics, University of Bergen, \\
All\'{e}gaten 55, N-5007 Bergen, Norway\\
$^{b}$Department of Physics, University of Texas, \\ 
Arlington, TX 76019, USA\\
\end{center}
\vfill

\abstract{ 
We discuss photon Bremsstrahlung induced by virtual graviton exchange in
proton--proton interactions at hadronic colliders, resulting from the exchange
of Kaluza--Klein excitations of the graviton. The relevant subprocesses,
$gg\to G\to e^+e^-\gamma$ and $q \bar q\to e^+e^-\gamma$ are discussed in
both the ADD and the RS scenarios. Although two-body final states (or real
graviton emission) would presumably be the main discovery channels, a search
for three-body final states could be worthwhile since such events have
characteristic features that could provide additional confirmation. In
particular, the $k_\perp$-distribution of the photon is in both scenarios 
harder than that of the Standard-model background.}
\vspace*{20mm} 
\setcounter{footnote}{0} 
\vfill

\newpage
\setcounter{footnote}{0}
\renewcommand{\thefootnote}{\arabic{footnote}}

\section{Introduction}\label{sec:I}
\setcounter{equation}{0}

The idea of additional compact dimensions and strings at the TeV scale,
proposed by Antoniadis \cite{Antoniadis:1990ew} for solving the hierarchy
problem, together with the idea that Standard-model (SM) fields live on branes
in a higher-dimensional space \cite{Akama:jy} have led to the even more
radical speculations that extra dimensions might be macroscopic, with SM
fields confined to the familiar four-dimensional world (brane)
\cite{Arkani-Hamed:1998rs,Randall:1999ee}. The models which allow for gravity
effects at the TeV scale can be grouped into two kinds: those of factorizable
geometry, where the extra dimensions are macroscopic
\cite{Arkani-Hamed:1998rs} (``ADD scenario''), and those of non-factorizable
(warped) geometry, the simplest example of which has only one extra dimension
separating ``our'' brane from a hidden brane \cite{Randall:1999ee} (``RS
scenario'').

In both these scenarios, the propagation of gravitons in the extra
dimensions leads to gravitons which from the four-dimensional
point of view are massive. In the ADD scenario, these Kaluza--Klein (KK)
gravitons have masses starting at values of the order of milli-eV, and there
is practically a continuum of them, up to some cut-off $M_S$ (close to 
the effective Planck scale)
of the order of TeV, whereas in the RS scenario they are
widely separated resonances with mass splittings of the order of TeV.
In both cases, they have a universal coupling to matter
and photons via the energy-momentum tensor.

These recent speculations have led to several studies 
\citer{Giudice:1999ck,Hewett:2002hv} of 
various experimental signals induced by graviton production and exchange.
The new scenarios allow for the emission of massive gravitons
\cite{Giudice:1999ck,Mirabelli:1999rt,Han:1999sg}, which would lead to 
events with missing energy (or transverse momentum), as well as 
effects due to the exchange of virtual gravitons (in addition to
photons and $Z$'s) 
\cite{Giudice:1999ck,Han:1999sg,Hewett:1999sn,Davoudiasl:2001wi}.
These processes include the production of dileptons and diphotons
in electron--positron collisions, 
as well as gluon--gluon and quark--antiquark-induced
processes at the Tevatron and LHC.

In fact, several searches at LEP and the Tevatron have given direct
bounds on the effective Planck scale, of the 
order of a TeV \citer{Hewett:2002hv,Acosta:2002eq}, while
astrophysical arguments result in very strong limits when applied to the
simplest ADD scenarios, for $n=2$ and 3 extra dimensions 
\cite{Hannestad:2001jv}.
Of course, the direct experimental searches are most worthwhile.
The above studies all focus on two-body final states,
which are expected to be dominant, and therefore lead to
the most stringent bounds on the existence of extra dimensions.

Here, we shall investigate photon Bremsstrahlung induced by graviton exchange 
\cite{Dvergsnes:2001fm}. While this cross section is further reduced by
${\cal O}(\alpha/\pi)$, so is the background.
It has some characteristic features resulting
from the exchange of a spin-2 particle and from the direct
graviton--photon coupling, that we would like to point out.
These features may be useful in discriminating any signal
against the background.

Specifically, we shall consider the process
\begin{equation}
\label{Eq:process-pp}
pp\to e^+e^-\gamma+X,
\end{equation}
which may get a contribution due to graviton exchange, and which
for energetic electrons (or muons) and photons should experimentally be 
a very clean signal.
(There is also a related process, where a graviton is emitted in the
final state \cite{Gabrielli:2002hn}.)

Since this final state is very distinct,
and since the Standard-model (Drell--Yan) background is well understood, 
the process (\ref{Eq:process-pp}) may offer some hope for observing 
a signal or improving on the exclusion bounds.

This paper is organized as follows: First (Sec.~\ref{sec:II}) we consider the
gluon--gluon fusion contribution to both two-body and three-body final
states. Then (Sec.~\ref{sec:III}) we consider quark--antiquark annihilation,
which also gives rise to the Standard-model background. We calculate the cross
section by summing over the KK tower within the ADD (Sec.~\ref{sec:IV}) and RS
(Sec.~\ref{sec:V}) scenarios for a selected choice of parameters, and finally
we give some concluding remarks (Sec.~\ref{sec:VI}).

\section{Gluon--gluon fusion}\label{sec:II}
\setcounter{equation}{0} 

We shall first discuss gluon--gluon fusion. Due to increasing gluon luminosity
at high energies (LHC), this contribution will be dominant for a certain range
in invariant mass of the (three-body) final state. At even higher invariant
masses, the contribution to the signal, from quark--antiquark annihilation,
becomes dominant.

\subsection{Two-body final states}

The process of interest, Eq.~(\ref{Eq:process-pp}), is related to
the two-body final state
\begin{equation}
\label{Eq:pp-ee}
pp\to e^+e^-+X,
\end{equation}
which may proceed via gluon--gluon fusion and an intermediate graviton,
\begin{equation}
gg\to G\to e^+e^-.
\end{equation}

For massless electrons, the cross section
for single graviton exchange resulting from gluon--gluon fusion is 
(in agreement with the results of \cite{Bijnens:2001gh})
\begin{equation}
\label{Eq:sigma-hat-ee}
\hat\sigma^{(G)}_{gg\to ee}
=\frac{\kappa^4 \hat s}{10240\pi}\,
\frac{\hat s^2}{(\hat s-m_{\vec n}^2)^2+(m_{\vec n}\Gamma_{\vec n})^2}, 
\end{equation}
with $\hat s=(k_1+k_2)^2$ the two-gluon invariant mass squared.
Furthermore, $m_{\vec n}$ and $\Gamma_{\vec n}$ are the mass and width of the 
graviton\footnote{The Kaluza--Klein index, ${\vec n}$ on 
$m_{\vec n}$ and 
$\Gamma_{\vec n}$ should not be confused with $n$, the number of extra 
dimensions.}, and $\kappa$ is the graviton coupling, to be defined below.
The angular distribution, which is forward--backward symmetric, is given by 
$1-\cos^4\theta$, where $\theta$ is the c.m.\ scattering angle.

With $\xi_1$ and $\xi_2$ the fractional momenta of the two gluons,
$k_1=\xi_1 P_1$, $k_2=\xi_2 P_2$, and $P_1$ and $P_2$ the proton 
momenta, $(P_1+P_2)^2=s$, we have $\hat s\simeq \xi_1\xi_2s$.
For the over-all process (\ref{Eq:pp-ee}) we thus find the differential cross
section 
\begin{eqnarray}
\label{Eq:dsigma-dshat}
\hspace{-0.7cm} \frac{d\sigma^{(G)}_{gg\to ee}}{d\hat s}
&\!\!\!=&\!\!\!\int_0^1d\xi_1\int_0^1d\xi_2\, f_g(\xi_1)f_g(\xi_2)\,
\frac{d\hat\sigma^{(G)}_{gg\to ee}}{d\hat s} \nonumber\\
\hspace{-0.7cm} &\!\!\!=&\!\!\!\int_0^1d\xi_1\int_0^1d\xi_2\, 
f_g(\xi_1)f_g(\xi_2)\, \delta\left(\xi_1\xi_2s-\hat s\right)\,
\hat\sigma^{(G)}_{gg\to ee}(\hat s) 
=\frac{1}{s}I_{gg}(\hat s)\, \hat\sigma^{(G)}_{gg\to ee}(\hat s),
\end{eqnarray}
with the relevant convolution integral, $I_{gg}(\hat s)$, over the gluon 
distribution functions given by Eq.~(\ref{Eq:I-convolutions}) in 
Appendix~A.

\subsection{Three-body final states}

Let us now consider the contribution from gluon--gluon fusion to the 
Bremsstrahlung process in Eq.~(\ref{Eq:process-pp}).
The underlying subprocess,
\begin{equation}
gg\to G\to e^+e^-\gamma,
\end{equation}
can proceed via the four Feynman diagrams of Fig.~\ref{Fig:gg-Feynman},
the basic couplings for which are given by Han et al.~\cite{Han:1999sg}
(see also Giudice et al.~\cite{Giudice:1999ck}).
\begin{figure}[htb]
\refstepcounter{figure}
\label{Fig:gg-Feynman}
\addtocounter{figure}{-1}
\begin{center}
\setlength{\unitlength}{1cm}
\begin{picture}(12,8)
\put(0.5,0.5)
{\mbox{\epsfysize=7.7cm\epsffile{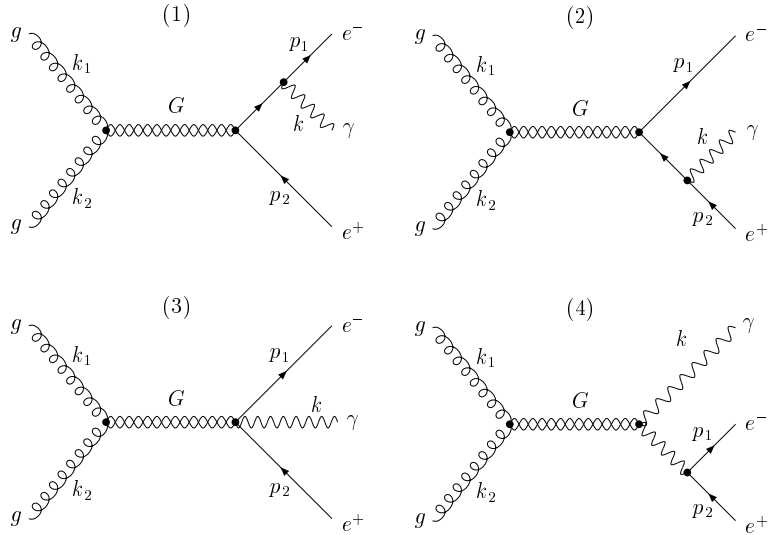}}}
\end{picture}
\caption{Feynman diagrams for $gg\to G\to e^+e^-\gamma$.  (There is a
corresponding set of diagrams with $q\bar q$ initial states, which we shall
refer to as ``set $B$''.  The SM diagrams corresponding to (1) and (2), with
$q\bar q$ initial states and a photon or $Z$ instead of the graviton, will be
referred to as ``set $D_\gamma$'' and ``set $D_Z$''.)}
\end{center}
\end{figure}

The evaluation of the cross section is straightforward, and the differential
cross section (w.r.t. the azimuthal angle, $\chi$, and $\cos \theta$, where
$\theta$ is the angle between the photon and the beam) is of fourth order in
the invariants. This is due to the
underlying mechanism being the exchange of a spin-2 object.
It is straight-forward to verify that it is gauge invariant with
respect to the gluons, as well as to the photon (actually, diagram 4 is by
itself gauge invariant).
But the expression is quite lengthy, so we shall not write it out here.

The angular distribution of the (two-body) non-radiative cross section 
(\ref{Eq:sigma-hat-ee}) is given by fourth-order polynomials in $\cos\theta$.
Here, just like in gluon Bremsstrahlung (see, e.g., \cite{Olsen:1980cw}), 
there is an accompanying dependence on the azimuthal angle $\chi$, 
but now up to fourth order in $\cos\chi$, or, equivalently, 
up to $\cos4\chi$.

After integrating over the azimuthal orientation of the events,
the cross section is of the form
\begin{equation}
\label{Eq:dsigma-cost}
\frac{d^3\hat\sigma^{(G)}_{gg\to ee \gamma}}{dx_1 dx_2 d\cos\theta}
\sim b_0(x_1,x_2)+b_2(x_1,x_2)\cos^2\theta+b_4(x_1,x_2)\cos^4\theta,
\end{equation}
similar to the two-body final states, i.e., the gluon--gluon fusion does not
contribute to any forward--backward asymmetry.

In our calculations we have chosen the unitary gauge ($\xi^{-1}=0$ in the
notation of \cite{Han:1999sg}), whereby the scalar exchange decouples. 
After averaging and summing over gluon, electron and photon polarizations,
and integrating over event orientations w.r.t.\ the gluon momentum, 
we find (for exchange of a single graviton)
\begin{equation}
\label{Eq:dsigma-gg-eega}
\frac{d^2\hat\sigma^{(G)}_{gg\to ee \gamma}}{dx_1dx_2}
=\frac{\alpha \kappa^4 \hat s Q_e^2}{2560\pi^2}\,
\frac{\hat s^2}{(\hat s-m_{\vec n}^2)^2+(m_{\vec n}\Gamma_{\vec n})^2}\,
\, X_B(x_1,x_2),
\end{equation}
with $X_B(x_1,x_2)$ given by Eq.~(\ref{Eq:XB}) in Appendix~A.
Furthermore, $\alpha$ is the fine-structure constant,
$Q_e=-1$ is the electron charge, and
$x_1$, $x_2$ and $x_3$ denote the fractional energies of the electrons
and the photon in the c.m.\ frame,
\begin{equation}
x_1=E_1/\sqrt{\hat s}, \qquad x_2=E_2/\sqrt{\hat s}, \qquad
x_3=\omega/\sqrt{\hat s}, \qquad 0\le x_i\le\half,
\end{equation}
with $x_1+x_2+x_3=1$. The expression $X_B(x_1,x_2)$
exhibits the familiar singularities in the infrared and collinear limits,
$s_1\equiv(p_1+k)^2=\hat s(1-2x_2)\to0$, $s_2\equiv(p_2+k)^2=\hat
s(1-2x_1)\to0$, as well as a collinear singularity at
$s_3\equiv(p_1+p_2)^2=\hat s(1-2x_3)\to0$ due to the fourth Feynman diagram.
Here $\hat s\equiv (k_1+k_2)^2=(p_1+p_2+k)^2$. The additional singularity
means that there is a tendency to have events with hard photons
\cite{Dvergsnes:2001fm}. This is one way in which these events differ from
ordinary QED-based Bremsstrahlung.

The differential cross section in Eq.~(\ref{Eq:dsigma-gg-eega}) can be written 
more compactly as
\begin{equation}
\frac{1}{\hat\sigma^{(G)}_{gg\to ee}}\,
\frac{d^2\hat\sigma^{(G)}_{gg\to ee \gamma}}{dx_1dx_2}
=\frac{4\alpha Q_e^2}{\pi}\, X_B(x_1,x_2),
\end{equation}
with $\hat\sigma^{(G)}_{gg\to ee}$ given by 
Eq.~(\ref{Eq:sigma-hat-ee}). As we see, the cross section is reduced by a
factor $\cal{O}(\alpha/\pi)$ compared to the two-body cross section.

Analogous to Eq.~(\ref{Eq:dsigma-dshat}), we find for the gluon contribution
to the over-all process (\ref{Eq:process-pp})
\begin{equation}
\label{Eq:dsigma-pp}
\frac{d^3\sigma^{(G)}_{gg\to ee \gamma}}{d\hat s\, dx_1dx_2}
=\frac{1}{s} I_{gg}(\hat s)\, 
\frac{d^2\hat\sigma^{(G)}_{gg\to ee \gamma}}{dx_1dx_2},
\end{equation}
with the convolution integral given by 
Eq.~(\ref{Eq:I-convolutions}) in Appendix~A.

\section{Quark--antiquark annihilation}\label{sec:III}
\setcounter{equation}{0}

Another process which contributes to (\ref{Eq:process-pp}), and gives rise to
the SM background, is quark--antiquark
annihilation. Furthermore, at large invariant masses of the final state, it
also gives the most important contribution to the signal. 

\subsection{Two-body final states}
The process in Eq.~(\ref{Eq:pp-ee}) may also proceed via quark--antiquark
annihilation and an intermediate graviton, with the following cross section for
single graviton exchange 
(initial state quarks are considered massless)
\begin{equation}
\label{Eq:qq-G-ee}
\hat{\sigma}^{(G)}_{q\bar q\to ee}
=\frac{\kappa^4\, \hat{s}}{15360 \pi}\,
\frac{\hat{s}^2}{(\hat{s}-m_{\vec n}^2)^2+m_{\vec n}^2\Gamma_{\vec n}^2},
\end{equation}
in agreement with \cite{Giudice:1999ck,Bijnens:2001gh}. It differs from the
cross section for gluon--gluon fusion by a factor $2/3$. 

There is also a SM background to this process, where the same final state is
produced through photon or $Z$ exchange. This well-known cross section is
given by
\begin{equation}
\label{Eq:sigma-qq-SM}
\hat{\sigma}^{(\text{SM})}_{q\bar q\to ee}(\hat s)
= \frac{4\pi \alpha^2}{9 \hat{s}}
\left[Q_q^2\, Q_e^2 +2 Q_q\,Q_e v_q v_e\, \Re\, \chi(\hat s)
+ (v_q^2 + a_q^2)(v_e^2 + a_e^2)|\chi(\hat s)|^2 \right], 
\end{equation}
with
\begin{equation}
\label{Eq:chi}
\chi(\hat s)=\frac{1}{\sin^2(2\theta_W)}\,
\frac{\hat s}{(\hat s-m_Z^2)+im_Z\Gamma_Z},
\end{equation}
where we have normalized vector and axial-vector couplings to 
$v_f=T_f-2Q_f\sin^2\theta_W$ and 
$a_f=T_f$ respectively, with $T_f$ the isospin. Furthermore, $m_Z$ and
$\Gamma_Z$ are the mass and width of the $Z$ boson, $Q_q$ the quark charge, and
$\theta_W$ the weak mixing angle.

In the case of $q\bar q \to G \to e^+e^-$, 
with the cross section given by Eq.~(\ref{Eq:qq-G-ee}), the angular 
distribution is $1-3\cos^2\theta+4\cos^4\theta$, 
whereas for the photon exchange, 
Eq.~(\ref{Eq:sigma-qq-SM}), the angular distribution is 
given by the familiar $1+\cos^2\theta$. The interference between
graviton and photon exchange has an angular distribution
given by $\cos^3\theta$ (as pointed out by
Ref.~\cite{Giudice:1999ck}) i.e., {\it it exhibits
a forward--backward asymmetry} and vanishes upon integration. The interference 
between graviton and $Z$ exchange exhibits a slightly different angular 
distribution (which also vanishes upon integration).

For $pp$ collisions, we find the graviton contribution to the differential 
cross section (in accordance with Eq.~(\ref{Eq:dsigma-dshat}))
\begin{equation}
\label{Eq:dsigma-dshat-qq}
\frac{d\sigma^{(G)}_{q \bar q \to ee}}{d\hat s}
=\frac{1}{s}I_{BB}(\hat s)\, 
\hat\sigma^{(G)}_{q \bar q \to ee}(\hat s),
\end{equation}
with $I_{BB}(\hat s)$ given in Appendix~A. The SM contribution can be
found in a similar manner, but here the convolution integrals must be weighted
by the factors $Q_q^2$, $Q_q v_q$ and $(v_q^2 + a_q^2)$ for photon exchange,
interference between the photon and the $Z$, and for $Z$ exchange, respectively
(see Appendix~A). The reason for this is that the convolution integral
implicitly contains flavor summation.

\subsection{Three-body final states}
Now we will examine the subprocess
\begin{equation}
q \bar q \to e^+e^-\gamma,
\end{equation}
which is determined by four sets of Feynman diagrams.  First, there are the
diagrams of Fig.~\ref{Fig:qq-Feynman}, referred to as ``set $A$'', describing
``initial-state radiation''. Then, there are four diagrams analogous to those
of Fig.~\ref{Fig:gg-Feynman}, where the initial-state gluons are replaced by
quarks and antiquarks. We shall refer to these as ``set $B$'', they describe
``final-state radiation''.  Finally, we have the SM background, which arises
from diagrams similar to 1 and 2 of sets $A$ and $B$, with a $\gamma$ or $Z$
exchanged instead of the graviton. We shall refer to the SM diagrams as ``sets
$C_\gamma$'', ``$C_Z$'' (initial-state radiation), ``$D_\gamma$'', and
``$D_Z$'' (final-state radiation).  It is convenient to separate initial-
from final-state radiation, since, in the former case, the propagator does not
carry all the momentum of the initial quarks.

\begin{figure}[htb]
\refstepcounter{figure}
\label{Fig:qq-Feynman}
\addtocounter{figure}{-1}
\begin{center}
\setlength{\unitlength}{1cm}
\begin{picture}(12,8)
\put(0.5,0.5)
{\mbox{\epsfysize=7.7cm\epsffile{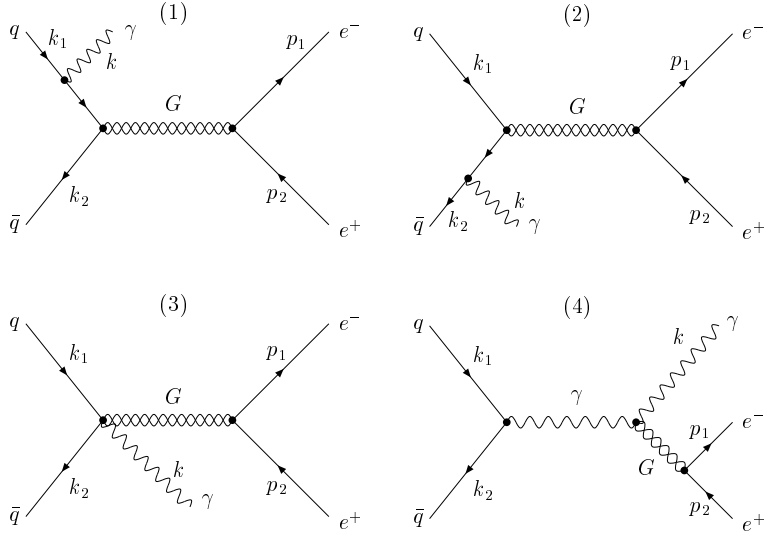}}}
\end{picture}
\caption{Feynman diagrams for ``initial-state radiation''
in $q\bar q\to e^+e^-\gamma$.
We refer to these diagrams as ``set $A$''.
The SM diagrams corresponding to (1) and (2), but with a photon or $Z$
instead of the graviton, are referred to as ``set $C_\gamma$''
and ``set $C_Z$''.}
\end{center}
\end{figure}

The $q\bar q$-initiated cross section can be decomposed as
\begin{equation}
\hat \sigma^{(G)}_{q\bar q\to ee \gamma}\, , \qquad
\hat \sigma^{(\text{SM})}_{q\bar q\to ee \gamma}\, , \qquad
\hat \sigma^{(G,\gamma)}_{q\bar q\to ee \gamma}\, , \qquad
\text{and} \qquad
\hat \sigma^{(G,Z)}_{q\bar q\to ee \gamma}\, ,
\end{equation}
where the first term is the graviton contribution (sets $A$ and $B$), the
second term is the Standard-model background (sets $C$ and $D$) and the last
two are graviton--photon and graviton--$Z$ interference terms, respectively.

First we shall consider the graviton exchange diagrams, where we introduce the
following notation,
\begin{equation}
\hat \sigma^{(G)}_{q\bar q\to ee \gamma}
=\hat \sigma_{A}+\hat \sigma_{B}
+\hat \sigma_{AB}.
\end{equation}

For initial-state radiation (set $A$), we find
\begin{equation}
\label{Eq:dsigma-qq-A}
\frac{d^2\hat\sigma_{A}}{dx_1dx_2}
=\frac{\alpha \kappa^4 s_3 \, Q_q^2 }{12288\pi^2}\,
\frac{s_3^2}{(s_3-m_{\vec n}^2)^2+(m_{\vec n}\Gamma_{\vec n})^2}\, 
X_A(x_1,x_2),
\end{equation}
where $X_A(x_1,x_2)$ is given by Eq.~(\ref{Eq:XA}) in Appendix~A.  Since the
denominator in Eq.~(\ref{Eq:dsigma-qq-A}) depends on $s_3=(1-2x_3)\hat s$
instead of $\hat s$, this contribution will be smeared out when integrated
over $x_3$.

The corresponding result for final-state radiation (set $B$), is
\begin{equation}
\frac{d^2\hat\sigma_{B}}{dx_1dx_2}
=\frac{\alpha \kappa^4 \hat s\, Q_e^2}{3840\pi^2}\,
\frac{\hat s^2}{(\hat s-m_{\vec n}^2)^2+(m_{\vec n}\Gamma_{\vec n})^2}\,
X_B(x_1,x_2),
\end{equation}
with $X_B(x_1,x_2)$ given by Eq.~(\ref{Eq:XB}). This contribution is, like in
the two-body case, identical to the contribution from gluon--gluon fusion,
except for a factor of $2/3$.

There is also an interference term, $\hat \sigma_{AB}$, between
sets $A$ and $B$, which has to be considered. It contributes to the
forward--backward asymmetry, and vanishes when integrated over all event
orientations.

In the three-body case, the SM background becomes
\begin{equation}
\hat \sigma^{(\text{SM})}_{q\bar q\to ee \gamma}
=\hat \sigma_{C}+\hat \sigma_{D}
+\hat \sigma_{CD}.
\end{equation}
where the contribution of initial- and final-state radiation is given by
\begin{align}
\label{Eq:dsigma-qq-SM}
\frac{d^2\hat\sigma_{C}}{dx_1dx_2}
&=\frac{3 \alpha Q_q^2}{4\pi}\, 
\hat\sigma^{(\text{SM})}_{q\bar q\to ee}(s_3)\,
X_{C}(x_1,x_2), \nonumber \\
\frac{d^2\hat\sigma_{D}}{dx_1dx_2}
&=\frac{4 \alpha Q_e^2}{\pi}\, 
\hat\sigma^{(\text{SM})}_{q\bar q\to ee}(\hat s)\,
X_{D}(x_1,x_2),
\end{align}
with $X_{C}(x_1,x_2)$ and $X_{D}(x_1,x_2)$ given in Appendix~A, and
$\hat\sigma^{(\text{SM})}_{q\bar q\to ee}$ given by
Eq.~(\ref{Eq:sigma-qq-SM}).  This $\hat\sigma_{D}$ is the familiar
Bremsstrahlung cross section expressed in terms of the related two-body
process.  While the contribution of the photon-exchange part of the
interference between initial- and final-state radiation to the integrated
cross section, $\hat\sigma_{C_\gamma D_\gamma}$, vanishes
\cite{Berends:1980yz}, this is not the case for terms involving $Z$-exchange
\cite{Berends:ie}.  They are included in the full SM background,
Eq.~(\ref{Eq:dsigma-sm}).

For the interference terms between graviton exchange and the SM diagrams,
we introduce the following notation:
\begin{align}
\hat \sigma^{(G,\gamma)}_{q\bar q\to ee \gamma}
&=\hat \sigma_{AC_\gamma}+\hat \sigma_{AD_\gamma} 
 +\hat \sigma_{BC_\gamma}+\hat \sigma_{BD_\gamma}, 
\nonumber \\[4pt]
\hat \sigma^{(G,Z)}_{q\bar q\to ee \gamma}
&=\hat \sigma_{AC_Z}+\hat \sigma_{AD_Z}
 +\hat \sigma_{BC_Z}+\hat \sigma_{BD_Z},
\end{align}
where the subscripts indicate the diagram sets involved.

We find that both $\hat \sigma_{AC_\gamma}$ and $\hat \sigma_{BD_\gamma}$,
together with the terms of $\hat \sigma_{AC_Z}$ and $\hat \sigma_{BD_Z}$
which are proportional to the vector coupling,
vanish after integration over event orientations, but they contribute to the
forward--backward asymmetry.  The $\hat \sigma_{AC_Z}$ term proportional to the
axial coupling also vanishes upon integration over all event configurations.
Finally, after integration over angles the $\hat \sigma_{BD_Z}$ term
proportional to the axial coupling, being proportional to the Legendre 
polynomial $P_2(\cos\theta)$, also vanishes.
(When introducing cuts, the cancellation is not complete.)

For the non-vanishing graviton-SM interference terms, we get
\begin{align}
\label{Eq:dsigma-qq-interference}
\frac{d^2\hat \sigma_{AD_\gamma}}{dx_1dx_2}
&=-\frac{\alpha^2 \kappa^2 Q_q^2 Q_e^2}{144 \pi}
\Re\left[\frac{s_3}{s_3-m_{\vec n}^2+i m_{\vec n}\Gamma_{\vec n}}\right]
X_{AD}(x_1,x_2), \nonumber \\
\frac{d^2\hat \sigma_{AD_Z}}{dx_1dx_2}
&= -\frac{\alpha^2 \kappa^2 Q_q Q_e v_q v_e}{144 \pi}
\Re\left[\chi^*(\hat s)\, \frac{s_3}{s_3-m_{\vec n}^2
+i m_{\vec n}\Gamma_{\vec n}}\right] X_{AD}(x_1,x_2),\nonumber \\
\frac{d^2\hat \sigma_{BC_\gamma}}{dx_1dx_2}
&=-\frac{\alpha^2 \kappa^2 Q_q^2 Q_e^2}{144 \pi}
\Re\left[\frac{\hat s}{\hat s-m_{\vec n}^2+i m_{\vec n}\Gamma_{\vec n}}\right]
X_{BC}(x_1,x_2), \nonumber \\
\frac{d^2\hat \sigma_{BC_Z}}{dx_1dx_2}
&= -\frac{\alpha^2 \kappa^2 Q_q Q_e v_q v_e}{144 \pi}
\Re\left[\chi^*(s_3)\, \frac{\hat s}{\hat s-m_{\vec n}^2
+i m_{\vec n}\Gamma_{\vec n}}\right] X_{BC}(x_1,x_2),
\end{align}
where $\chi(\hat s)$ is given by Eq.~(\ref{Eq:chi}), whereas $X_{AD}(x_1,x_2)$
and $X_{BC}(x_1,x_2)$ are given in Appendix~A.

To find the contribution from quark--antiquark annihilation to the over-all
process~(\ref{Eq:process-pp}), a relation similar to the one given by
Eq.~(\ref{Eq:dsigma-dshat-qq}) should be used. 
However, when there are quark charges or vector/axial-vector couplings
($v_q$ or $a_q$) involved, the convolution integrals must be weighted 
by these factors, as shown in Appendix~A.

\section{Bremsstrahlung in the ADD scenario}\label{sec:IV}
\setcounter{equation}{0}
In this section we shall consider Bremsstrahlung within the ADD scenario
\cite{Arkani-Hamed:1998rs}. First, we shall present the cross section as a
function of invariant mass, and then study the photon distribution (or
$k_\perp$ spectrum) of such events.

\subsection{Total cross section}\label{sec:add-total}
In the ADD scenario, 
the coupling of each KK mode to matter is Planck-scale suppressed.
However, since the states are very closely spaced, with \cite{Han:1999sg}
\begin{equation} 
m_{\vec n}^2=\frac{4\pi^2\vec n^2}{R^2},
\end{equation}
and $R/2\pi$ the compactification radii,
the coherent summation over the many modes leads to effective couplings
with strength $1/M_S$, where $M_S$ is the UV cut-off (close to the effective
Planck scale).

Explicitly, in this scenario, the graviton coupling
is in the $(4+n)$-dimensional theory given by \cite{Han:1999sg}
\begin{equation}
\hat g_{MN}=\hat \eta_{MN}+\hat \kappa \hat h_{MN}, \qquad 
\hat \kappa^2=16\pi G_{\rm N}^{(4+n)},
\end{equation}
where $G_{\rm N}^{(4+n)}$ is Newton's constant in $4+n$ dimensions.
In $4$ dimensions the coupling can be written as
\begin{equation}
\kappa^2=V_n^{-1} \hat \kappa^2=16\pi V_n^{-1}\,G_{\rm N}^{(4+n)}
=16\pi G_{\rm N},
\end{equation}
with $V_n$ the volume of the $n$-dimensional compactified space
($V_n=R^n$ for a torus $T^n$) and $G_{\rm N}$ the 4-dimensional 
Newton constant.

Summing coherently over all KK modes in a tower, the graviton propagator 
gets replaced \cite{Han:1999sg}:
\begin{equation}
\label{Eq:propagatorsum}
\frac{1}{\hat s-m_{\vec n}^2+im_{\vec n} \Gamma_{\vec n}} 
\equiv -iD(\hat s)
\to \frac{\hat s^{n/2-1}}{\Gamma(n/2)}\,
\frac{R^n}{(4\pi)^{n/2}}\, [2I(M_S/\sqrt{\hat s}) -i\pi],
\end{equation}
with 
\begin{equation}
\label{Eq:III}
I(M_S/\sqrt{\hat s})=
\begin{cases}
\begin{displaystyle}
-\sum_{k=1}^{n/2-1}\frac{1}{2k}\, \left(\frac{M_S}{\sqrt{\hat s}}\right)^{2k} 
- \frac{1}{2}\log\left(\frac{M_S^2}{\hat s} - 1\right), 
\end{displaystyle}
& n=\text{even}, \\
\begin{displaystyle}
- \sum_{k=1}^{(n-1)/2}\frac{1}{2k-1} 
\left(\frac{M_S}{\sqrt{\hat s}}\right)^{2k-1} 
+ \frac{1}{2}\log\left(\frac{M_S + \sqrt{\hat s}}{M_S - \sqrt{\hat s}}\right), 
\end{displaystyle}
& n=\text{odd}.
\end{cases}
\end{equation}
for $n$ extra dimensions. 

Higher order loop effects may be important \cite{Contino:2001nj}, so these
expressions should not be taken too literally. In particular, this applies to
the dependence on the number of extra dimensions. In the approach of
\cite{Giudice:1999ck} and \cite{Hewett:1999sn} this uncertainty, including the
$n$-dependence, is absorbed in the cut-off in such a way that $D(\hat s)$ and
$D(s_3)$ (see Eq.~(\ref{Eq:propagatorsum})) are
indistinguishable.  Here, in order to preserve the qualitative difference
between these two propagators, related to final- and initial-state radiation,
we shall use the expressions of Eq.~(\ref{Eq:III}).

Invoking the relation \cite{Han:1999sg}
\begin{equation}
\kappa^2 R^n =8\pi(4\pi)^{n/2}\Gamma(n/2)M_S^{-(n+2)}
\end{equation}
between the volume of the extra dimensions, the gravitational coupling
and the cut-off scale, the differential cross section can be expressed 
as
\begin{align}
\label{Eq:dsigma-all-add}
\frac{d^3\sigma}{d\hat s\, dx_1dx_2} 
&=\frac{\alpha\, s_3}{192 s M _S^{4}}\, 
\left(\frac{s_3}{M _S^2}\right)^{n}\, 
I_{D_\gamma D_\gamma}(\hat s) \left[4I^2(M_S/\sqrt{s_3})+\pi^2 \right]
X_A(x_1,x_2) \nonumber \\ 
&+\frac{\alpha\, Q_e^2 \hat s}{120 s M_S^4}\,
\left(\frac{\hat s}{M _S^2}\right)^{n} 
[3 I_{gg}(\hat s) + 2 I_{BB}(\hat s)] 
\left[4I^2(M_S/\sqrt{\hat s})+\pi^2 \right] 
X_B(x_1,x_2) \nonumber \\
&-\frac{ \alpha^2}{18 s M_S^{2}}\,
\bigg[
\left(\frac{s_3}{M _S^2}\right)^{n/2}
\Big(2Q_e^2 I_{D_\gamma D_\gamma}(\hat s)\, I(M_S/\sqrt{s_3})\, 
\nonumber \\ 
&\phantom{MMMMMM}
+Q_e v_e I_{D_\gamma D_Z}(\hat s)\, 
\Re \big\{\chi^\ast(\hat s)\big[2I(M_S/\sqrt{s_3})-i\pi\big]\big\}
\Big) X_{AD}(x_1,x_2) \nonumber \\
&\phantom{MMlMM}
+\left(\frac{\hat s}{M _S^2}\right)^{n/2}
\Big(2Q_e^2 I_{D_\gamma D_\gamma}(\hat s)\, I(M_S/\sqrt{\hat s})\, 
\nonumber \\ 
&\phantom{MMMMMM}
+Q_e v_e I_{D_\gamma D_Z}(\hat s)\, 
\Re \big\{\chi^\ast(s_3)\big[2I(M_S/\sqrt{\hat s})-i\pi\big]\big\}
\Big)
X_{BC}(x_1,x_2)\bigg] \nonumber \\ 
&+\frac{d^3\sigma^{\rm (SM)}}{d\hat s\, dx_1dx_2},
\end{align}
where $\chi(\hat s)$ is given by Eq.~(\ref{Eq:chi}) and the convolution
integrals are given in Appendix~A. 
The SM background is given by
\begin{align}
\label{Eq:dsigma-sm}
\frac{d^3\sigma^{\rm (SM)}}{d\hat s\, dx_1dx_2} 
&=\frac{\alpha^3}{3 s s_3}
\big\{Q_e^2\, I_{C_\gamma C_\gamma}(\hat s)
+2 Q_e v_e\, I_{C_\gamma C_Z}(\hat s)\,\Re\, \chi(s_3) \nonumber \\
&\phantom{MMMMM}
+ (v_e^2 + a_e^2) I_{C_Z C_Z}(\hat s)|\chi( s_3)|^2
\big\}\, X_{C}(x_1,x_2)  \nonumber \\
&+\frac{16 \alpha^3 Q_e^2}{9 s \hat s}
\big\{ Q_e^2 I_{D_\gamma D_\gamma}(\hat s) 
+2 Q_e v_e I_{D_\gamma D_Z}(\hat s)\, 
\Re\, \chi(\hat s) \nonumber \\
&\phantom{MMMMM}
+ (v_e^2 + a_e^2) I_{D_Z D_Z}(\hat s)|\chi(\hat s)|^2 \big\}
\, X_{D}(x_1,x_2) \nonumber \\
&-\frac{8 \alpha^3 Q_e a_e}{3 s \hat s}
\big\{Q_e\,I_{C_\gamma D_Z}(\hat s) \,\Re\, \chi(\hat s)
+Q_e I_{C_Z D_\gamma}(\hat s) \,\Re\, \chi(s_3) \nonumber \\
&\phantom{MMMMM}
+4 v_e I_{C_Z D_Z}(\hat s) \,\Re[\chi^\ast(\hat s)\chi(s_3)] \big\}
\, X_{CD}(x_1,x_2).
\end{align}
In Eq.~(\ref{Eq:dsigma-all-add}), the different contributions are given
in the following order: First quark--antiquark annihilation with initial-state
radiation (set $A$), then gluon--gluon fusion and quark--antiquark
annihilation with final-state radiation (set $B$), various interference terms
between graviton exchange and SM amplitudes, and finally the SM
background. The origins of the different terms are reflected in the subscripts
of the $X$'s.
\begin{figure}[htb]
\refstepcounter{figure}
\label{Fig:add}
\addtocounter{figure}{-1}
\begin{center}
\setlength{\unitlength}{1cm}
\begin{picture}(16.2,7.3)
\put(0.0,0.0)
{\mbox{\epsfysize=8.0cm\epsffile{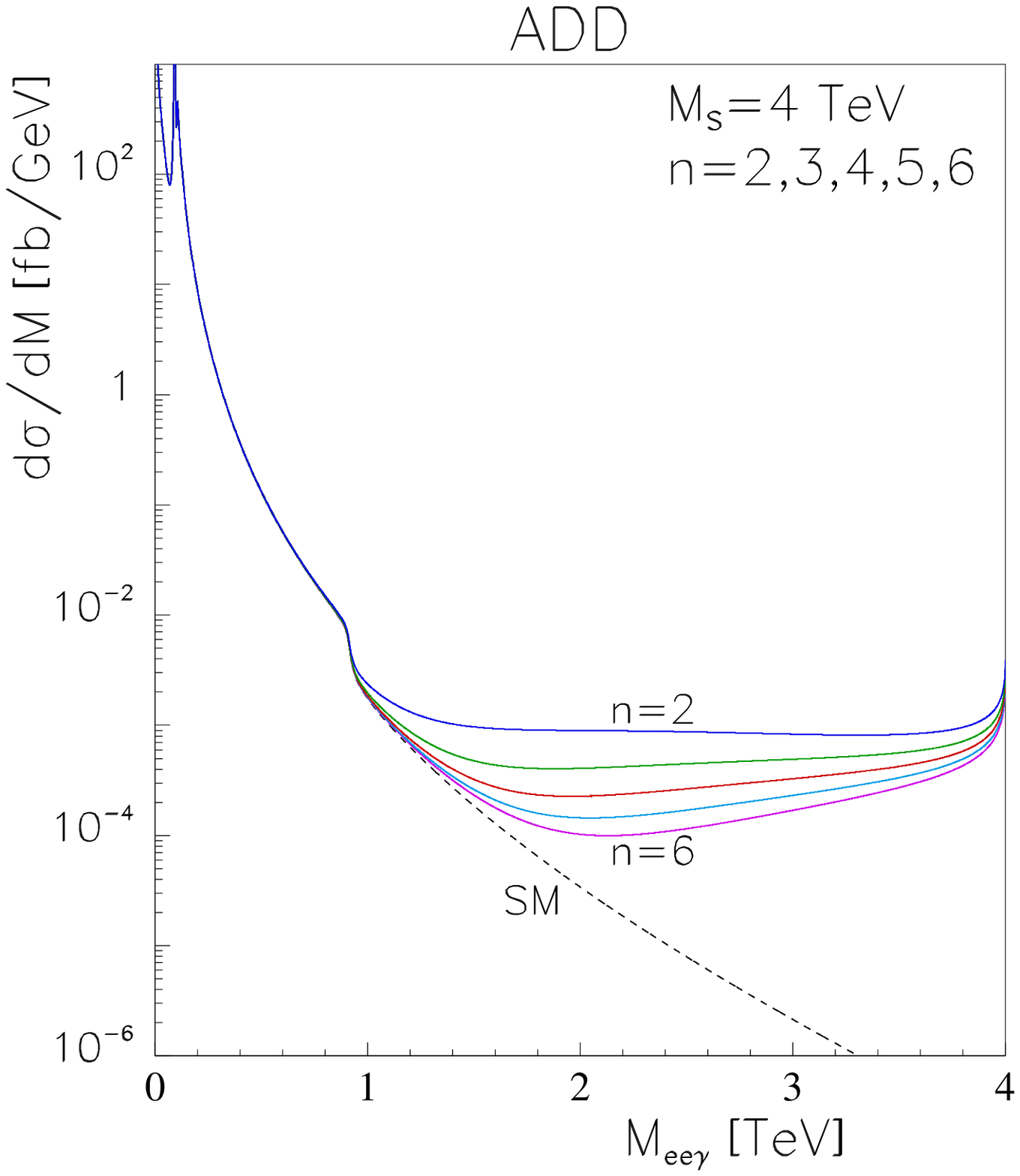}}
 \mbox{\epsfysize=8.0cm\epsffile{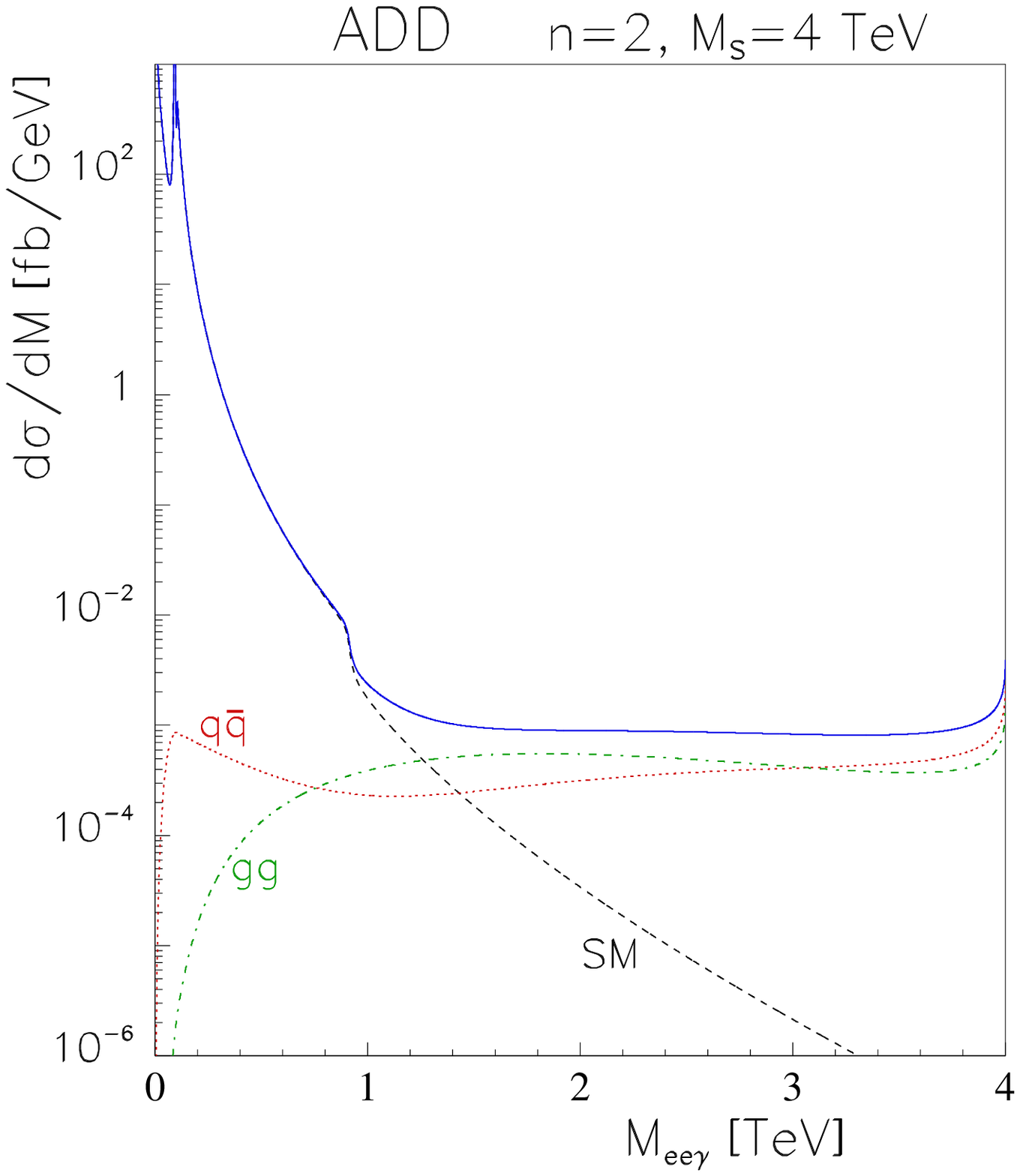}}}
\end{picture}
\caption{Cross sections for $pp\to e^+e^-\gamma +X$ at
$\sqrt{s}=14~\text{TeV}$.  Both plots are for the ADD model with
$M_S=4$~TeV. We have set the number of extra dimensions (from above) to
$n=2,3,4,5,$ and $6$ for the left panel.  The full differential cross section,
$d\sigma/d\sqrt{\hat s}$ (solid), and the SM background (dashed) are shown in
both plots, whereas the $gg$ (dash-dotted) and the $q \bar q$ (dotted)
contributions to the signal are shown in the right panel for $n=2$.}
\end{center}
\end{figure}

In Fig.~\ref{Fig:add}, we show the cross section, differential w.r.t.\
$\sqrt{\hat s}$ (labeled $M_{ee\gamma}$ in the figures), for $M_S=4$~TeV and
$n=2,3,4,5$ and $6$ (left panel) in the ADD scenario, where we have integrated
out the $x_1,x_2$ dependence (see Appendix~B). The right panel shows only the
$n=2$ curve, with the contributions from gluon--gluon fusion and
quark--antiquark annihilation (induced by graviton exchange) also displayed.
Note that gluon--gluon fusion is dominant from $\sqrt{\hat s}\simeq 1.3$~TeV up
to $\sqrt{\hat s}\simeq 3$~TeV for this choice of parameters, whereas the
quark--antiquark annihilation process takes over at larger invariant masses.
Below $\sim1$~TeV, the background is larger than the signal.

We have integrated over $x_3^{\rm min}\le x_3 \le 0.5$, subject to the
$y$-cuts: $s_1, s_2 \ge y \hat s$, $s_3\ge y_3 \hat s$, where both
$y=y_3=0.01$.  The minimum invariant mass of the two electrons is controlled
by $y_3$.  At a scale $\sqrt{\hat s}=1$~TeV, the cut of $y=0.01$ corresponds
to electron (or photon) energies exceeding 10~GeV.  We consider a minimum
$x_3$ of $0.1$ in these plots.  The corresponding resolution
is well within that foreseen at the LHC \cite{lhc}.
For the angular cuts, we take the pseudorapidity $|\eta|<2.5$.

Here, and in the remaining figures, we use $\sqrt{s}=14$~TeV, which
corresponds to the LHC energy. With an integrated luminosity of
$100~\text{fb}^{-1}$ and a bin-width of $100$~GeV, we might expect a few
events per bin at invariant masses above $1$~TeV.

Near the $Z$ mass, the cross section has an additional peak (barely visible in
the figures) since both the amplitudes with $\hat s$ (set $C_Z$) and those
with $s_3$ (set $D_Z$) resonate.  Also, at $\hat s\sim 10\times m_Z\sim
0.9\text{ TeV}$, the cross section has a fall-off.  In fact, both the second
peak and the fall-off are related to ``radiative return'', where the emitted
photon has the right energy to make the $Z$ propagator resonate.  The peak
near $m_Z$ is at the starting point for radiative return, determined by $s_3$
(and hence the lower value for $x_3$), whereas the fall-off near $0.9\text{
TeV}$ is related to the end point of the radiative return, given by the upper
value of $x_3$, which is $\half(1-y_3^{\rm cut})$.

Close to the cut-off, $M_S$, the cross section blows up due to the logarithm
in $I(M_S/\sqrt{\hat s})$. This is of course an artifact, due to the
way the cut-off is treated \cite{Contino:2001nj}.
Note that the explicit $n$-dependence in Fig.~\ref{Fig:add} is in the 
approach of \cite{Giudice:1999ck,Hewett:1999sn} absorbed in the cut-off.

Let us now discuss the different terms related to $q\bar q$ annihilation.  The
part of the $q\bar{q}$ annihilation cross section which is related to $X_A$
(initial-state radiation with graviton exchange) is everywhere significantly
smaller than the one related to $X_B$ (final-state radiation with graviton
exchange), by more than an order of magnitude. This is partly due to the
difference in the convolution integrals.  

Among the different interference terms, related to $X_{AD}$ and $X_{BC}$, the
most important one is that between set~$B$ (final-state radiation with
graviton exchange) and set~$C_\gamma$ (initial-state radiation with photon
exchange).  For invariant masses below $\sim1\text{ TeV}$, this interference
dominates the $q\bar q$ part of the signal (but here, it is overwhelmed by the
SM background).  We note that the SM amplitude does not interfere with the
gluon--gluon fusion part of the graviton-mediated amplitude.
\subsection{Photon distribution}
Whereas photons emitted by QED Bremsstrahlung tend to be soft and/or collinear
with the fermions, the present graviton-induced Bremssstrahlung will be more
energetic \cite{Dvergsnes:2001fm}, and also emitted at larger angles.  A
first, qualitative manifestation of this feature is that the transverse
momentum spectrum (with respect to the incident beam) will be harder.  We
shall here make this statement quantitative.

In Sec.~\ref{sec:add-total}, we have integrated the differential cross section
over $x_1,x_2$ and $\cos\theta$ to obtain cross sections which are
differential w.r.t.\ $\sqrt{\hat s}$ (see also
Eq.~(\ref{Eq:dsigma-dshat})). Now we will instead change variables from
$(x_1,x_2,\cos\theta)$ via $(x_{12}=x_1-x_2,x_3,\cos\theta)$ to
$(x_{12},k_\perp,k_\|)$, where $k_\perp=\omega\sin\theta$ and
$k_\|=\omega\cos\theta$.  Here
\begin{equation}
\sqrt{k_\perp^2+k_{\|}^2}=k=\omega=x_3\sqrt{\hat s},
\end{equation}
with $\omega$ the energy of the photon.
After integration over $x_{12}$, $k_\|$ and $\sqrt{\hat s}$, we 
get the cross section, differential w.r.t.\ $k_\perp$, the photon momentum
perpendicular to the beam.

\begin{figure}[htb]
\refstepcounter{figure}
\label{Fig:add-kt}
\addtocounter{figure}{-1}
\begin{center}
\setlength{\unitlength}{1cm}
\begin{picture}(16.2,7.7)
\put(0.0,0.0)
{\mbox{\epsfysize=8.0cm\epsffile{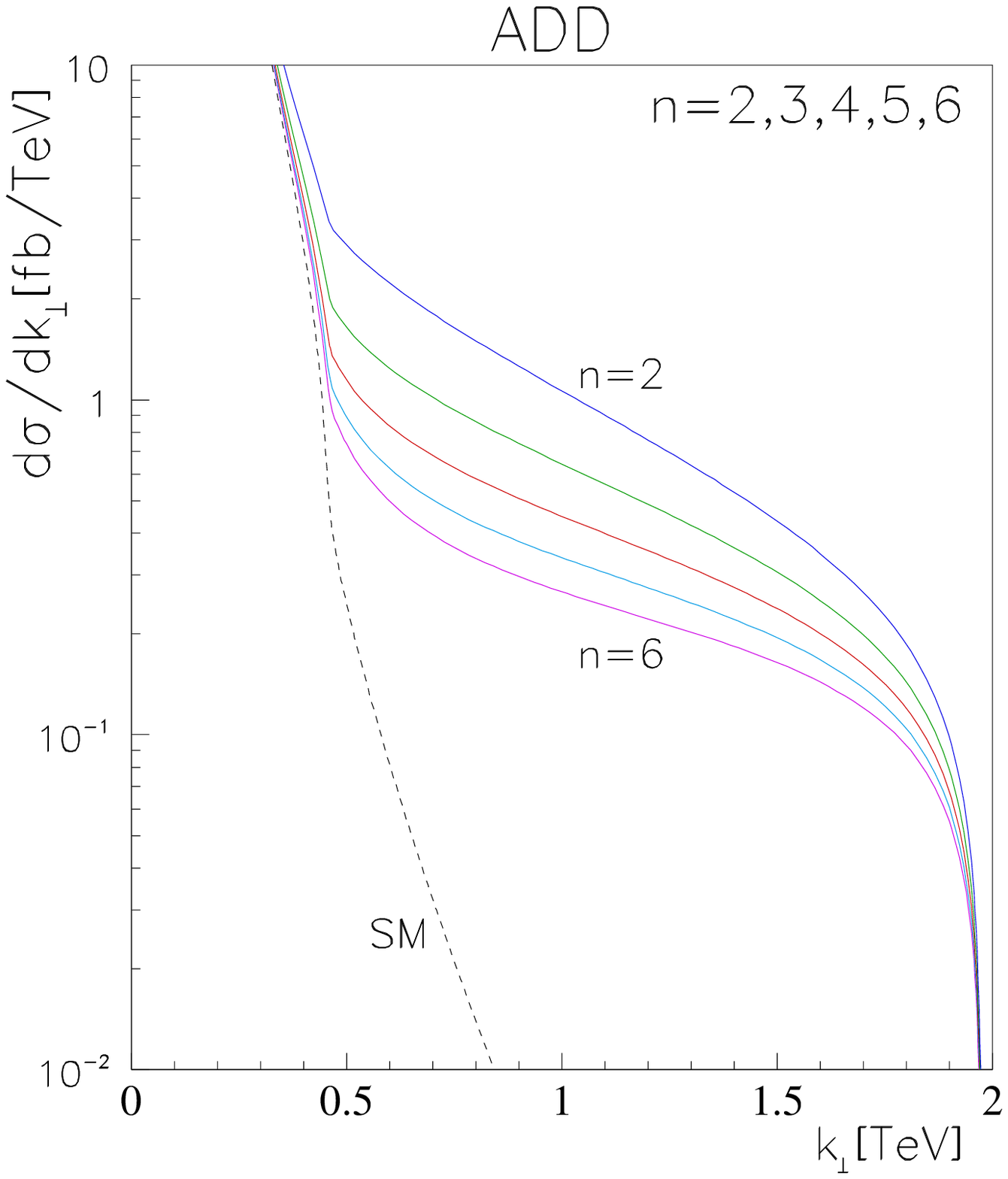}}
 \mbox{\epsfysize=8.0cm\epsffile{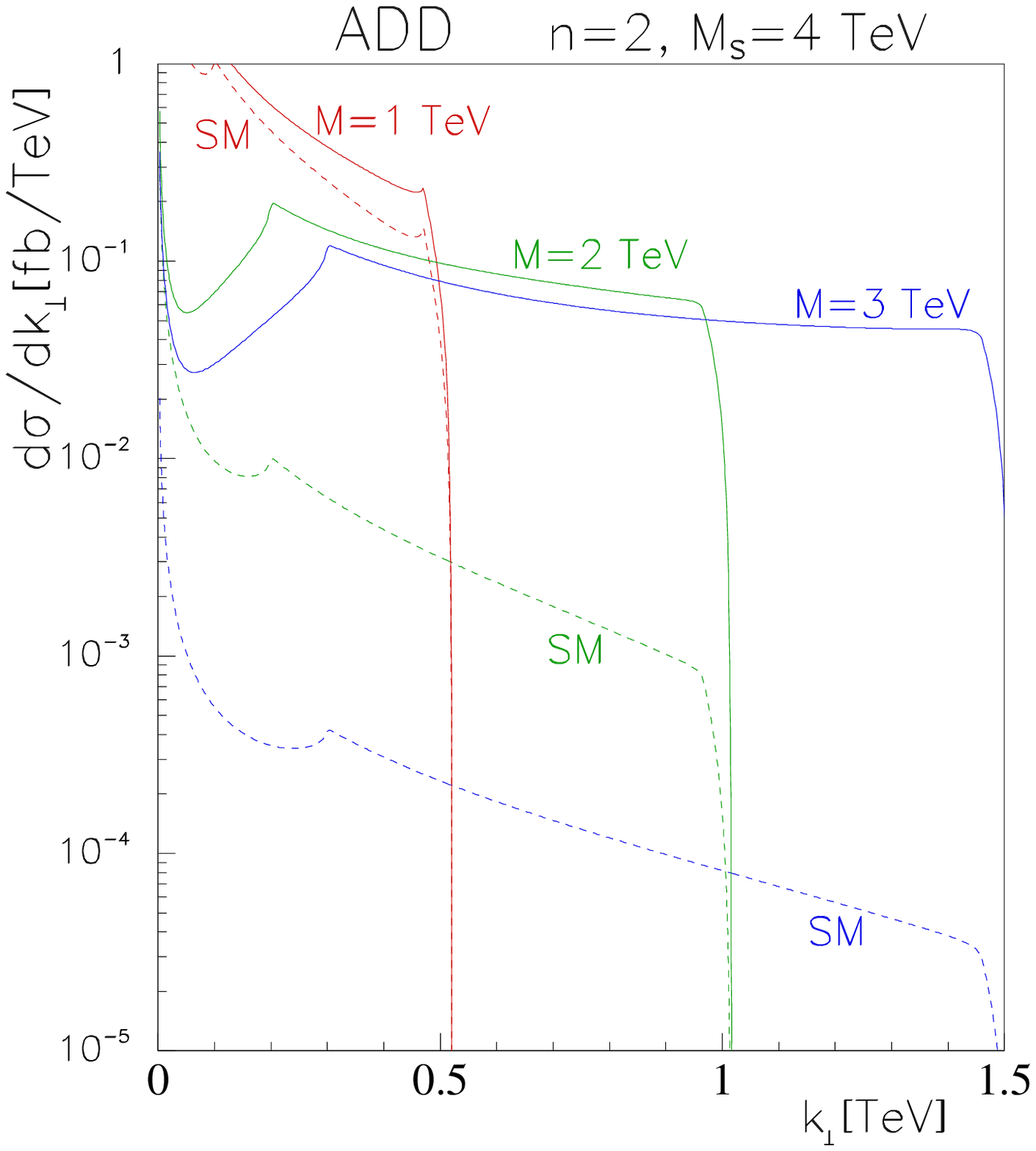}}}
\end{picture}
\caption{Photon distributions for the ADD model.
Left panel: $d\sigma/dk_\perp$ (solid) with (from above)
$n=2,3,4,5,6$ extra dimensions and $M_S=4$~TeV, together with the SM
background. 
Right panel: $d\sigma/dk_\perp$ (solid) for $n=2$ and $M_S=4$~TeV, integrated 
over $100$~GeV bins around $M_{ee\gamma}=M=1,2,3$~TeV, with the corresponding 
SM backgrounds (dashed).}
\end{center}
\end{figure}

In Figure~\ref{Fig:add-kt}, left panel, we show $d\sigma/dk_\perp$ for 
different numbers of extra dimensions, and with $M_S=4$~TeV, where we have
integrated over $\sqrt{\hat s}$ up to the cut-off, $M_S$. In accordance with
Fig.~\ref{Fig:add}, we impose the cuts: 
$x_3^{\rm cut}\sqrt{\hat s} \le k \le \half(1-y_3)\sqrt{\hat s}$. 
Since the photon should not be too close to the beam axis, we also require
$k_\perp \ge 100$~GeV. 
We see that the $k_\perp$-spectrum becomes much harder when
extra dimensions are involved.

The right panel of Fig.~\ref{Fig:add-kt} shows bin-integrated cross
sections. Here we have chosen invariant masses of $\sqrt{\hat s}=1,2,3$~TeV,
and integrated $d^2\sigma/(dk_\perp d\sqrt{\hat s})$ over $100$~GeV bins around
these values. The peaks at $0.1$, $0.2$ and $0.3$~TeV are also related to
cuts on $k$ (or $x_3$). Note that the SM background is very small
at higher invariant masses.

\section{Bremsstrahlung in the RS scenario}\label{sec:V}
\setcounter{equation}{0}
As in the previous section we will also for the Randall--Sundrum scenario 
\cite{Randall:1999ee}
first present the total cross section, and
thereafter the photon distribution.
\subsection{Total cross section}
In the RS scenario, 
the graviton masses are given by \cite{Davoudiasl:2000jd}
\begin{equation}
m_n=k x_n\, e^{-kr_c\pi} = \frac{x_n}{x_1} m_1,
\end{equation}
where $x_n$ are roots of the Bessel function\footnote{The
first four roots are $3.83,\ 7.02,\ 10.17$ and $13.32$.} of order 1, 
$J_1(x_n)=0$,
$k$ is of the order of the (four-dimensional) Planck scale and $r_c$ 
the compactification radius of the extra 
dimension\footnote{To solve the hierarchy problem, $kr_c \sim 12$ is 
required.}. Since there is only one extra dimension in this scenario, we shall
use $m_n$ instead of $m_{\vec n}$ for the mass of the $n$th graviton.

The gravitational coupling is in this model given 
by \cite{Davoudiasl:2000jd,Bijnens:2001gh}
\begin{equation}
\kappa=\sqrt{16 \pi}\,\frac{x_n}{m_n}\, \frac{k}{M_{\rm Pl}}
=\sqrt{2}\,\frac{x_1}{m_1}\, \frac{k}{\overline M_{\rm Pl}}, \qquad
\overline M_{\rm Pl}=\frac{M_{\rm Pl}}{\sqrt{8\pi}}
=2.4\times10^{18}~\text{GeV}.
\end{equation}
Here we shall use $m_1$ and $k/\overline M_{\rm Pl}$ as the two parameters
which specify the model. Note that $0.01 \leq k/\overline M_{\rm Pl} \leq 1$
\cite{Davoudiasl:2000jd}.

The differential cross section can in the RS scenario be expressed as
(with the different contributions given in the same order as in
Eq.~(\ref{Eq:dsigma-all-add})):
\begin{align}
\label{Eq:dsigma-all-rs}
\frac{d^3\sigma}{d\hat s\, dx_1dx_2} 
&=\frac{\alpha s_3}{3072 \pi^2 s}
\left(\frac{x_1}{m_1}\right)^4 
\left(\frac{k}{\overline M_{\rm Pl}}\right)^4 
I_{D_\gamma D_\gamma}(\hat s)
\left| {\cal G}(s_3)\right|^2 X_A(x_1,x_2) \nonumber \\
&+ 
\frac{\alpha \hat s}{1920 \pi^2 s}
\left(\frac{x_1}{m_1}\right)^4
\left(\frac{k}{\overline M_{\rm Pl}}\right)^4 
Q_e^2 
[3 I_{gg}(\hat s) + 2 I_{BB}(\hat s)]
\left| {\cal G}(\hat s)\right|^2
\, X_B(x_1,x_2) \nonumber \\
&-\frac{\alpha^2}{72 \pi s}
\left(\frac{x_1}{m_1}\right)^2
\left(\frac{k}{\overline M_{\rm Pl}}\right)^2 \nonumber \\
&\times
\Big[
\Big(Q_e^2 I_{D_\gamma D_\gamma}(\hat s)\,
\Re\, {\cal G}(s_3) 
+Q_e v_e I_{D_\gamma D_Z}(\hat s)\,
\Re \left[\chi^\ast(\hat s)\, {\cal G}(s_3)\right]
\Big)X_{AD}(x_1,x_2)\nonumber \\
&\quad
+\Big(Q_e^2\, I_{D_\gamma D_\gamma}(\hat s)\,
\Re \, {\cal G}(\hat s)
+Q_e v_e I_{D_\gamma D_Z}(\hat s)\,
\Re \left[\chi^\ast(s_3) {\cal G}(\hat s)\right]
\Big)X_{BC}(x_1,x_2)\Big] \nonumber \\
&+\frac{d^3\sigma^{\rm (SM)}}{d\hat s\, dx_1dx_2},
\end{align}
where the SM contribution is given by Eq.~(\ref{Eq:dsigma-sm}).  For massive
graviton exchange we have introduced the abbreviation
\begin{equation} \label{Eq:cal-gravity}
{\cal G}(\hat s)=\sum_{n} 
\frac{\hat s}{\hat s -m_n^2+im_n\Gamma_n}
\end{equation}
with \cite{Han:1999sg,Allanach:2000nr}
\begin{equation}
\label{Eq:Gammai}
\Gamma_n\equiv\frac{\gamma_G}{20\pi}\, m_n^3\kappa^2 
= \frac{\gamma_G}{10\pi}\, x_n^2 m_n 
\left(\frac{k}{\overline M_{\rm Pl}}\right)^2,
\end{equation}
and
\begin{equation}
\gamma_G=1+\chi_\gamma+\chi_Z+\chi_W+\chi_\ell+\chi_q+\chi_H+\chi_r
\end{equation}
the total graviton width in units of the two-gluon width.
Neglecting mass effects, we have \cite{Han:1999sg,Allanach:2000nr}
\begin{eqnarray}
\chi_\gamma&=&\frac{1}{8}, \qquad
\chi_Z=\frac{13}{96}, \qquad
\chi_W=\frac{13}{48}, \nonumber \\
\chi_\ell&=&\frac{N_\ell}{16}, \qquad
\chi_q=\frac{N_c N_q}{16}, \qquad
\chi_H=\chi_r=\frac{1}{48}.
\end{eqnarray}
Here, $N_\ell=6$ is the number of leptons, and $N_c N_q=18$ is the number of
quarks weighted with color factors. Note that since we have neglected quark
and electron masses, there is no contribution to the cross section from radion
exchange, since the radion couples to the trace of the energy-momentum
tensor. However, it contributes slightly to $\Gamma_n$ through
Eq.~(\ref{Eq:Gammai}).
\begin{figure}[htb]
\refstepcounter{figure}
\label{Fig:rs}
\addtocounter{figure}{-1}
\begin{center}
\setlength{\unitlength}{1cm}
\begin{picture}(16.2,7.3)
\put(0.0,0.0)
{\mbox{\epsfysize=8.0cm\epsffile{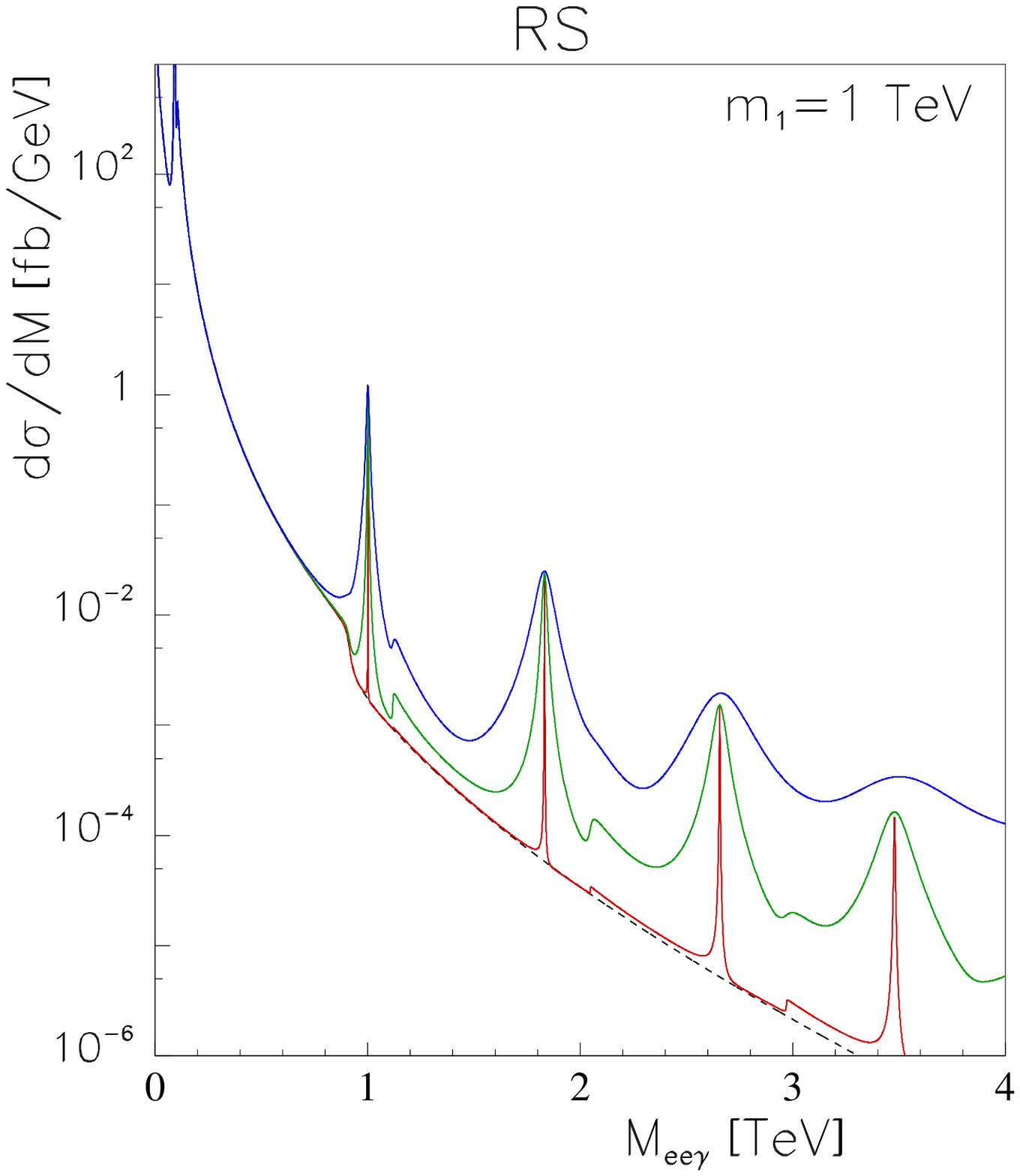}}
 \mbox{\epsfysize=8.0cm\epsffile{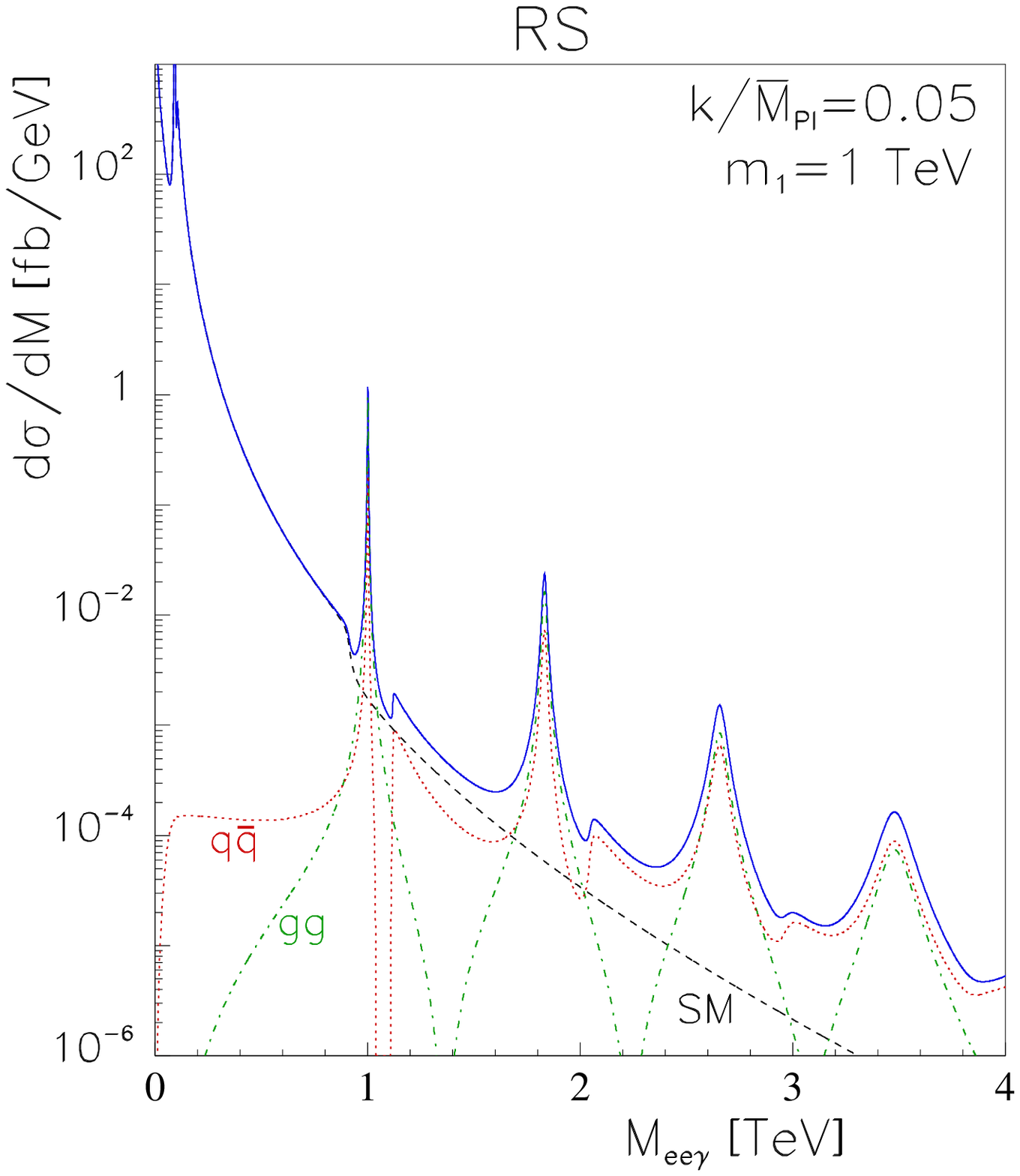}}}
\end{picture}
\caption{Differential cross sections for the RS model, with $m_1=1$ TeV.
The left panel shows the differential cross section, 
$d\sigma/d\sqrt{\hat s}$ (solid) with (from above)
$k/\overline M_{\rm Pl}=0.1$, $0.05$ and $0.01$, and the SM background
(dashed). Right panel: the $gg$ (dash dotted) and 
$q \bar q$ (dotted) contributions to the cross section 
are shown for $k/\overline M_{\rm Pl}=0.05$.}
\end{center}
\end{figure}

We display the RS scenario cross section, differential w.r.t.\ $\sqrt{\hat s}$
(see Eq.~(\ref{Eq:dsigma-all-rs})), in Fig.~\ref{Fig:rs}. In the left panel,
we have summed over KK states, and chosen the first graviton resonance at
$m_1=1$~TeV, with $k/\overline M_{\rm Pl}$ set to $0.1$, $0.05$ and $0.01$ 
(from above). 
In the right panel, we show the different contributions ($gg$ and $q \bar q$) 
to the cross section (for $k/\overline M_{\rm Pl}=0.05$) induced by graviton
exchange. The cuts are the same as in the ADD case. 

As we mentioned above, the integration over $s_3$ (or $x_3$) will smear out
the contribution from initial-state radiation. For invariant masses greater
than
\begin{equation} \label{Eq:rad-return}
\sqrt{\hat s}=\frac{m_1}{\sqrt{1-2x_3^{\rm min}}}\simeq1.1\; m_1,
\end{equation}  
$s_3^{\rm max}$ will always be greater than $m_1$, and radiative return to the
first graviton resonance gives a small peak at this value,
(\ref{Eq:rad-return}), and similarly for the higher resonances.  
In fact, these secondary peaks are mainly due to the term involving $X_A$
(initial-state bremsstrahlung with graviton exchange),
which gives no visible effect in the ADD case.

Similar to the ADD case, for invariant masses below $\sim1\text{ TeV}$,
it is the interference between set $B$ and set $C_\gamma$ which dominates
the $q\bar q$ signal (dotted). Between the peak at 1~TeV and the secondary
peak next to it (due to radiative return) this interference term
becomes negative, resulting in a dip in the total cross section.
In fact, the total $q\bar q$ signal (graviton exchange plus its interference
with the SM amplitude) is negative in some small region of invariant mass.
However, these structures depend on details of the cuts imposed.

It should be noted that the third and fourth peaks should be somewhat reduced 
since we have not taken into account that these gravitons can decay into the 
first KK resonance. Self-interactions of the gravitons were considered in
\cite{Davoudiasl:2001uj}, where a BR of about $15$\% was found for the $G_3
\to G_1 G_1$ decay.

By comparing Figs.~\ref{Fig:add} and~\ref{Fig:rs}, we see that the difference
between the two scenarios is striking, with extremely narrow, and widely
separated resonances in the RS case, compared to the continuum in the ADD
case. Note that according to our expressions, the $k/\overline M_{\rm Pl}$
dependence cancels at the resonance. A single peak should therefore have a
height which is independent of $k/\overline M_{\rm Pl}$, but at very high
invariant masses we see that the $k/\overline M_{\rm Pl}=0.1$ peak is higher
than the other peaks, and also slightly shifted. This is mainly due to
interference with, and overlap of the neighboring peaks which are very broad.

\begin{figure}[htb]
\refstepcounter{figure}
\label{Fig:rs-int-bin}
\addtocounter{figure}{-1}
\begin{center}
\setlength{\unitlength}{1cm}
\begin{picture}(16.2,14.8)
\put(0.0,7.6)
{\mbox{\epsfysize=8.0cm\epsffile{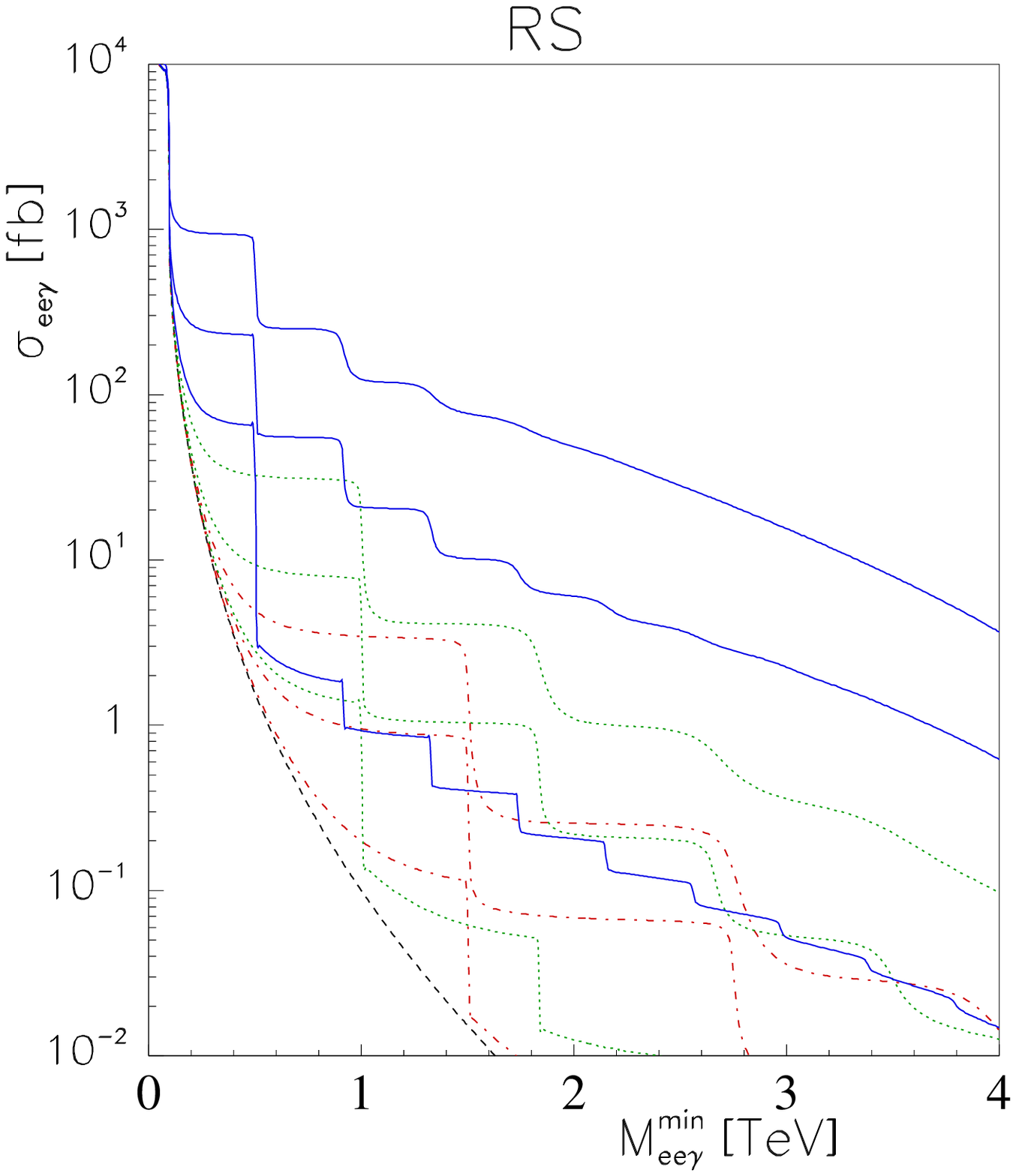}}
 \mbox{\epsfysize=8.0cm\epsffile{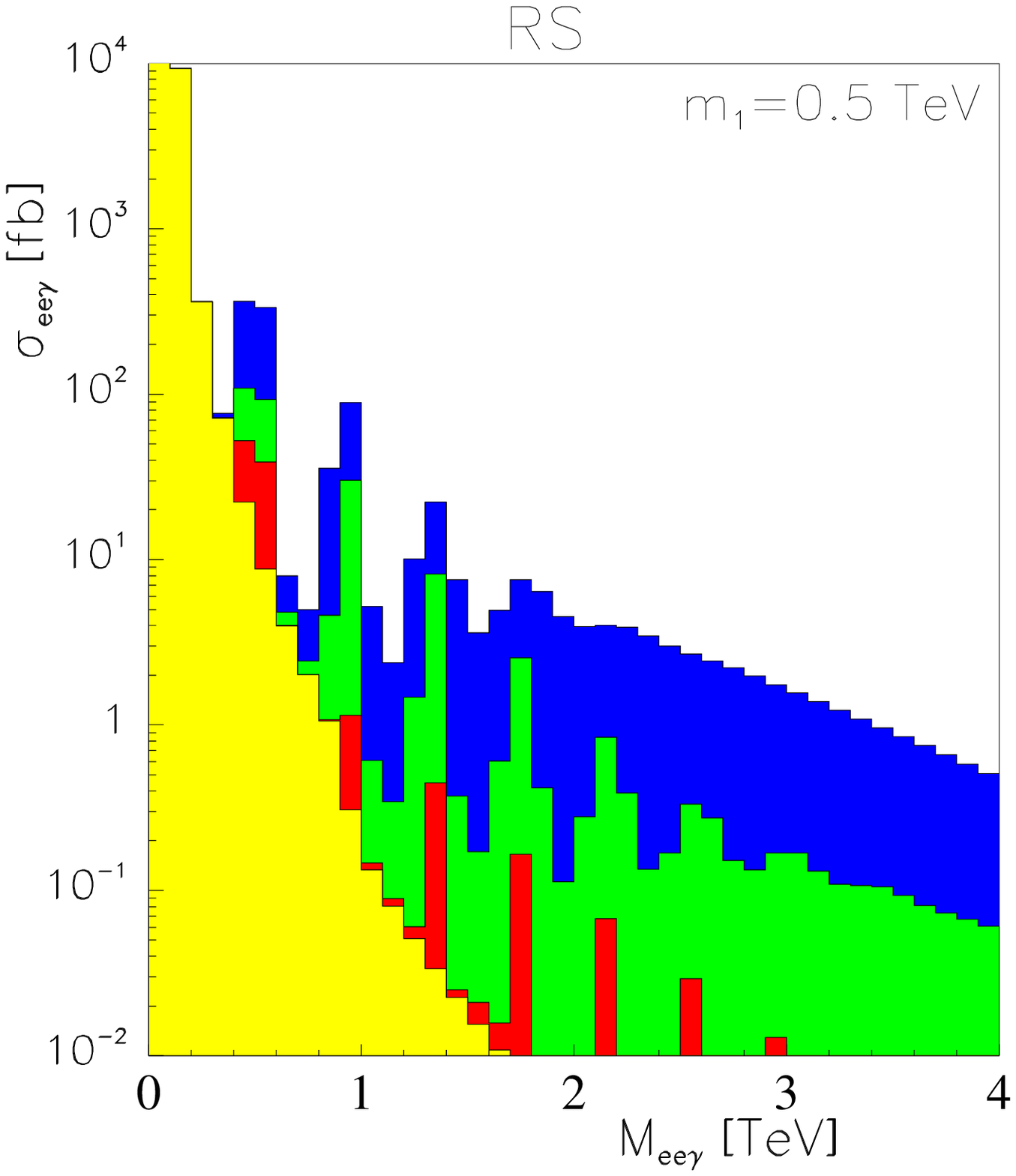}}}
\put(0.0,-0.4)
{\mbox{\epsfysize=8.0cm\epsffile{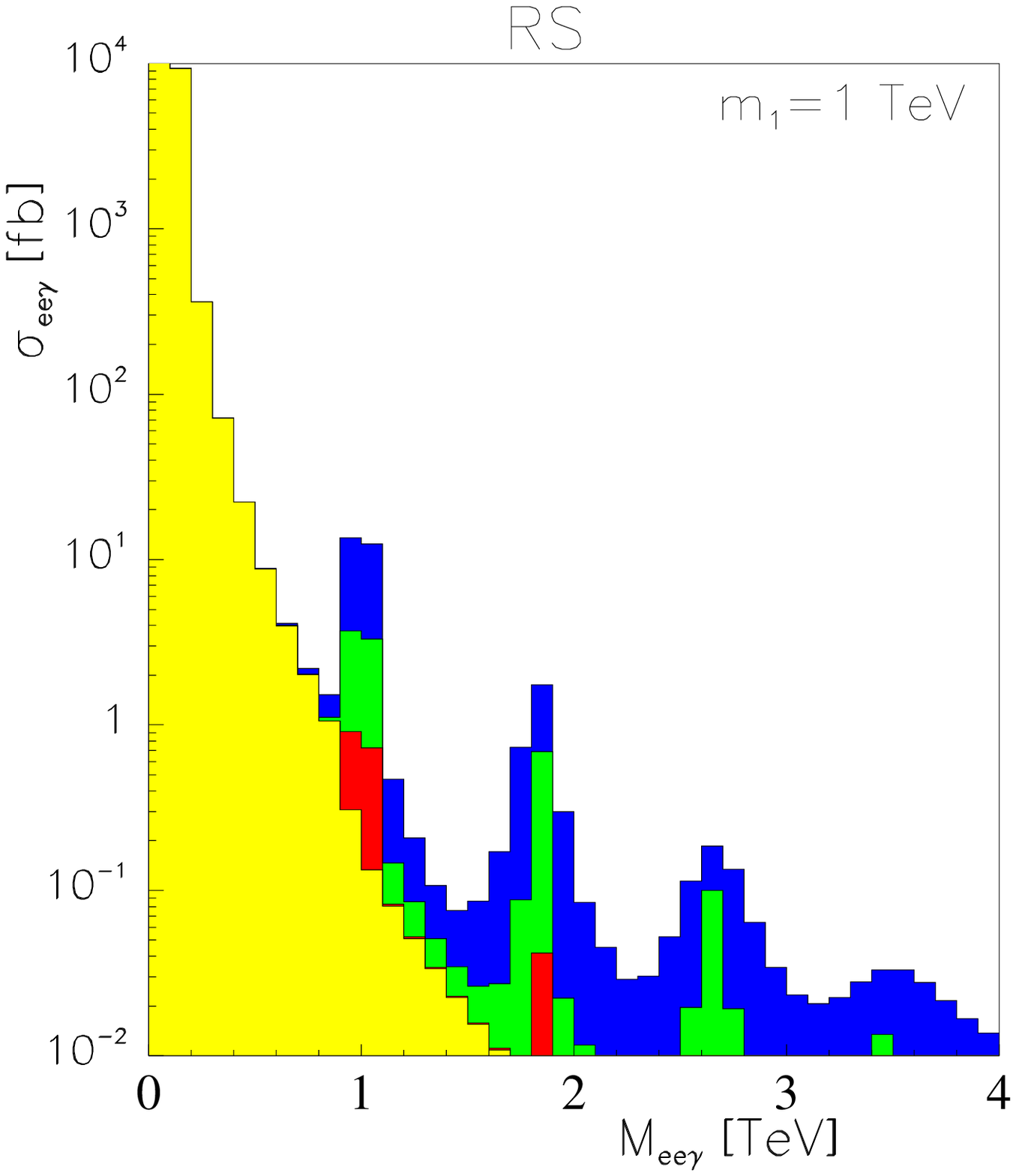}}
 \mbox{\epsfysize=8.0cm\epsffile{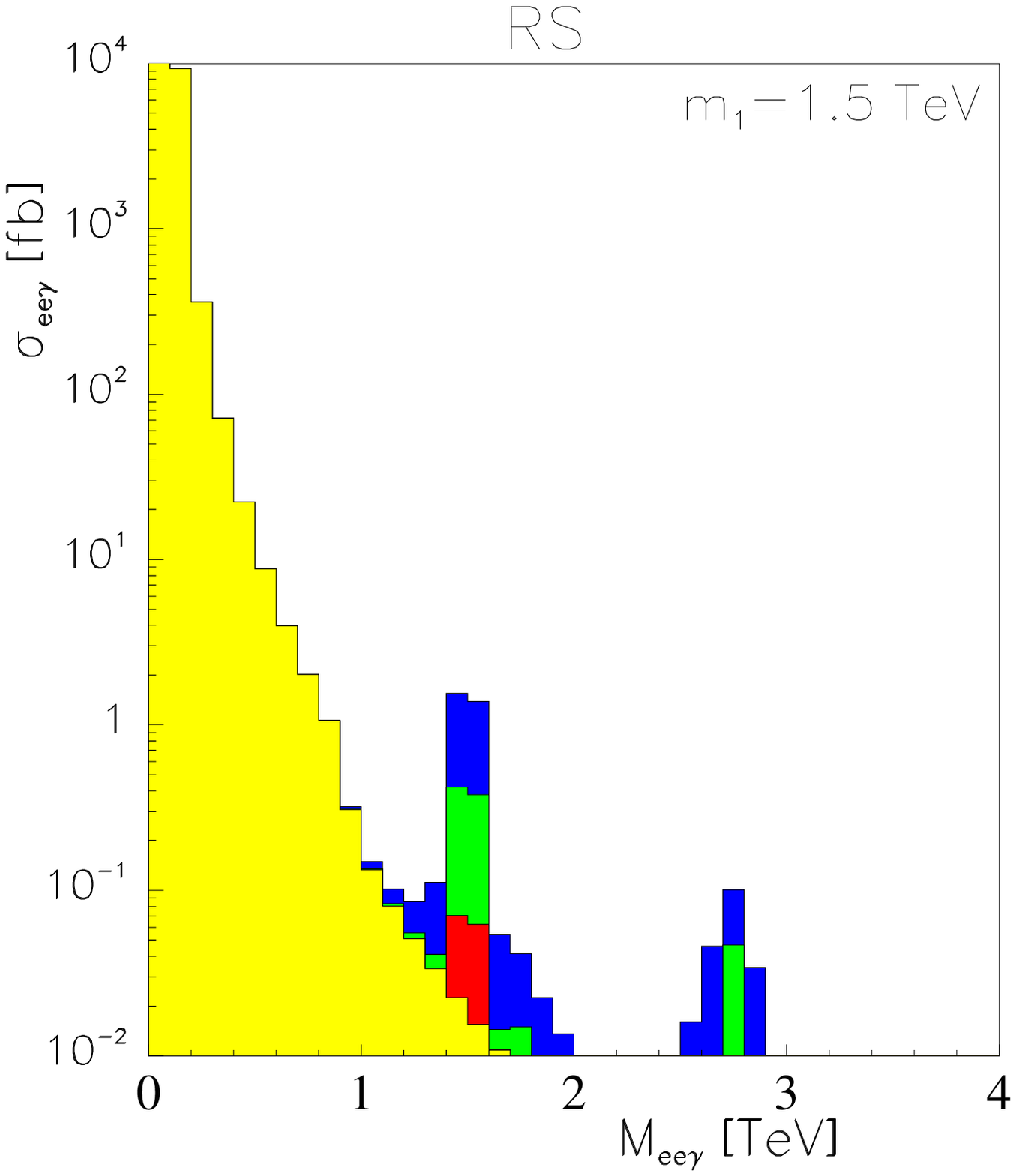}}}
\end{picture}
\caption{Integrated cross sections for the RS scenario.
In the upper left panel we have integrated the cross sections in
Fig.~\ref{Fig:rs} over $\sqrt{\hat s}$ for 
$\sqrt{\hat s}\ge M^{\rm min}_{ee\gamma}$. 
The choice of parameters is $k/\overline M_{\rm Pl}=0.1$, $0.05$ and
$0.01$ (from above) for $m_1=0.5$ (solid), $1$ (dotted) and $1.5$~TeV
(dash-dotted). The SM background (dashed) is also shown. The other three
panels show the corresponding result of integrating over $100$~GeV bins, 
with $m_1=0.5$ (upper right), $m_1=1$ (lower left) and $m_1=1.5$~TeV 
(lower right).}
\end{center}
\end{figure}

Since the cross sections for graviton-induced Bremsstrahlung are much lower
than for the corresponding two-body final state, an important question is,
however, if there is any chance of seeing these resonances in the experiments.
To give an order of magnitude estimate of the number of events to expect from
these narrow peaks, we have integrated the differential cross sections given
in Fig.~\ref{Fig:rs} over bins in $M_{ee\gamma}$. In the upper left panel of
Fig.~\ref{Fig:rs-int-bin} we integrate over 
$\sqrt{\hat s}$, starting from
$\sqrt{\hat s}=M_{ee\gamma}^{\rm min}$. The different curves correspond to
$k/\overline M_{\rm Pl}=0.1$, $0.05$ and 0.01 (from above) for $m_1=0.5$
(solid), $1$ (dotted) and $1.5$~TeV (dash-dotted). This should be compared to
the SM background (dashed) which is also shown. In the remaining panels of
Fig.~\ref{Fig:rs-int-bin} we have integrated over $100$~GeV bins for the same
choice of $k/\overline M_{\rm Pl}$, for $m_1=0.5$, 1.0 and 1.5~TeV.

We see that for these parameters it should be possible to resolve at least the
first peak, and in most cases several peaks are visible. We emphasize
that this is not a simulation, but these plots should provide an indication of
the number of events that correspond to these narrow peaks.

\subsection{Photon distribution}
The photon distribution is also in the RS case harder than the SM
background.  In the left panel of Fig.~\ref{Fig:rs-kt} we show
$d\sigma/dk_\perp$, having integrated over $\sqrt{\hat s}$. The steps occur
when $k_\perp$ equals half the invariant mass of a RS-graviton, since this
limits the photon momentum.

The right panel of Fig.~\ref{Fig:rs-kt} again shows bin-integrated cross
sections, but now we have chosen $100$~GeV bins around the first three
resonances, $\sqrt{\hat s}=m_1$, $m_2$, $m_3$, since this is where most of the
events will occur. We see a similar behaviour as in the ADD scenario.

\begin{figure}[htb]
\refstepcounter{figure}
\label{Fig:rs-kt}
\addtocounter{figure}{-1}
\begin{center}
\setlength{\unitlength}{1cm}
\begin{picture}(16.2,7.7)
\put(0.0,0.0)
{\mbox{\epsfysize=8.0cm\epsffile{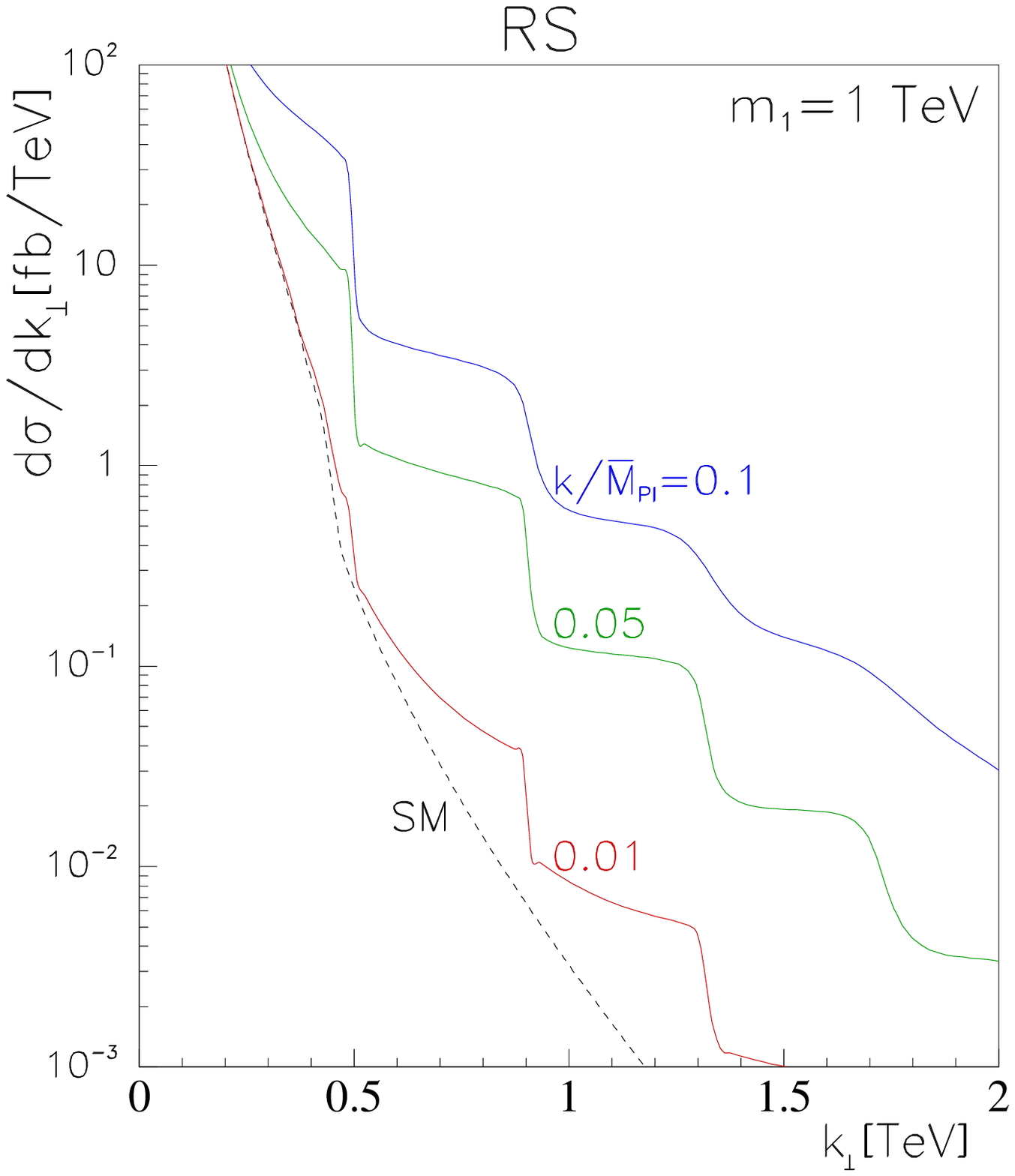}}
 \mbox{\epsfysize=8.0cm\epsffile{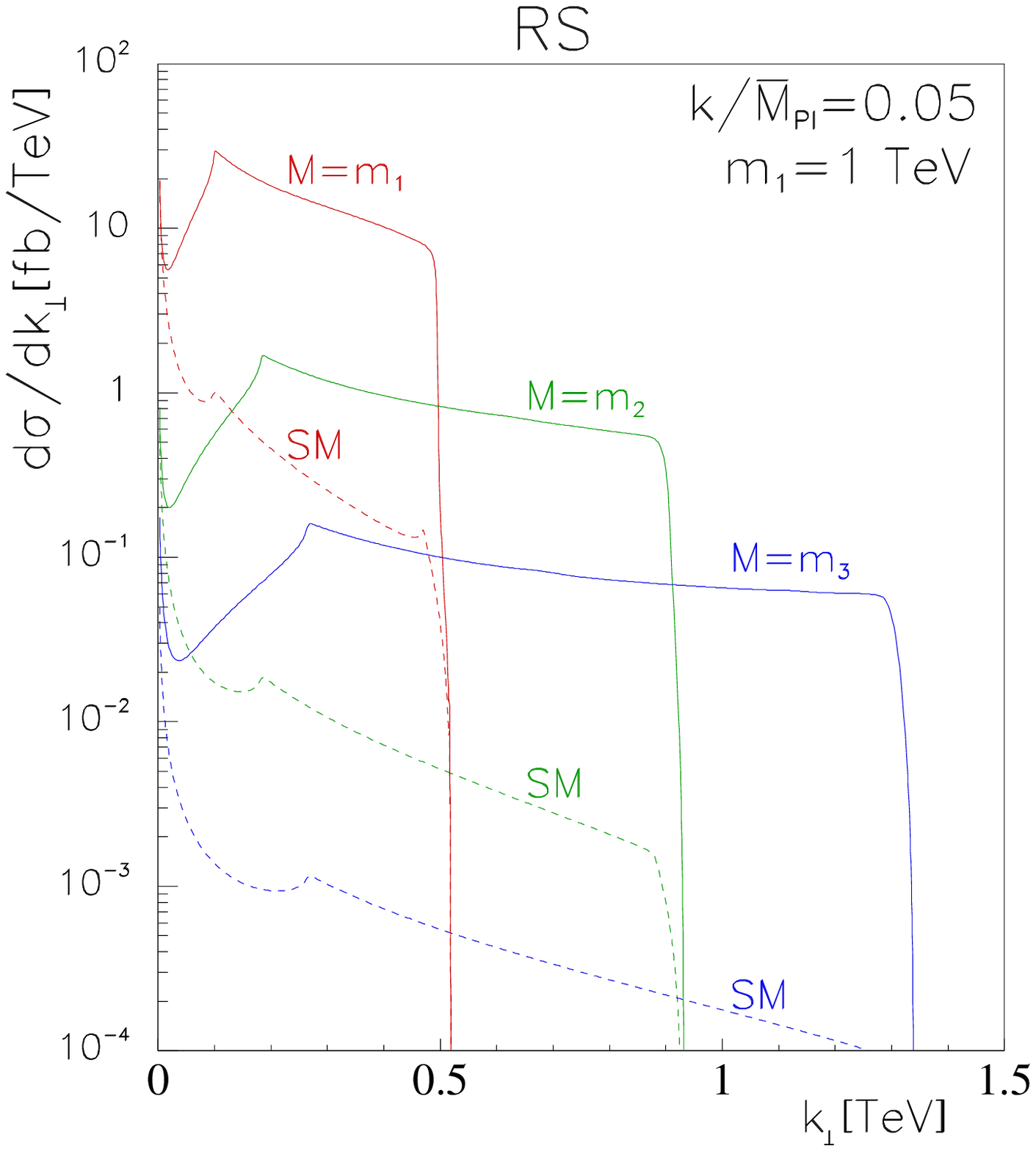}}}
\end{picture}
\caption{Photon distributions for the RS model.
Left panel: $d\sigma/dk_\perp$ (solid) with (from above)
$k/\overline M_{\rm Pl}=0.1$, $0.05$, $0.01$, $m_1=1$ and the SM background
(dashed). 
Right panel: $d\sigma/dk_\perp$ (solid) with $m_1=1$ and 
$k/\overline M_{\rm Pl}=0.05$, integrated over $100$~GeV bins around the first
three resonances, with corresponding SM backgrounds.}
\end{center}
\end{figure}

\section{Concluding remarks}\label{sec:VI}
\setcounter{equation}{0}

In summary, we have discussed photon Bremsstrahlung induced by the exchange of
massive gravitons at the LHC. Both the ADD and the RS scenarios have been
considered. We found that three-body final states are likely to be detectable,
and could be a valuable supplement to the two-body final states, for the
purpose of detecting the effects from massive gravitons related to extra
dimensions (if such exist). These configurations, of a hard photon associated
with an electron--positron (or muon) pair in the opposite direction, should
provide a striking signature at the LHC.  Furthermore, the photon has a harder
distribution in transverse momentum than the SM background.

We have here focused on Bremsstrahlung at the LHC. At the Fermilab, the
phenomenology will be rather different. Because of the lower energy, and
because of the different initial state, quark--antiquark annihilation will be
much more important, relative to the gluon--gluon initial state. Furthermore,
there will in $p\bar p$ collisions be additional contributions to the
forward--backward asymmetries, beyond those of the SM. Such asymmetries will
be induced by interference between the C-odd exchange of a photon or a $Z$,
and the C-even exchange of a graviton, as well as by the interference between
graviton-exchange with initial-state radiation and graviton-exchange with
final-state radiation. This effect, which is washed out at the LHC because of
the symmetry of the initial state, is of course present also for $e^+e^-\to f
\bar f$ \cite{Hewett:1999sn}, and will have an analogue in the three-body
final states. We hope to discuss this effect elsewhere.

\bigskip

{\bf Acknowledgments.}
It is a pleasure to thank Hans Bijnens, John Ellis, 
Gian Giudice, JoAnne Hewett, and, in particular,
Paolo di Vecchia for very useful discussions.
This research has been supported by the Research Council of Norway
and by NORDITA. The work of N.\"O. was supported in part by the
Robert A. Welch Foundation.

\clearpage

\appendix
\section*{Appendix A: Convolution Integrals and Event Shapes}
\setcounter{equation}{0}
\renewcommand{\thesection}{A}
In this Appendix, we define the convolution integrals and the event
shapes used in our expressions.
First, the basic convolution integrals are:
\begin{align}
\label{Eq:I-convolutions}
I_{gg}(\hat s) &= \int_{-Y}^{Y} dy\,
f_g\left(\sqrt{\frac{\hat s}{s}}\,e^y\right)
f_g\left(\sqrt{\frac{\hat s}{s}}\,e^{-y}\right), \nonumber \\
I_{q\bar q}(\hat s)&=\int_{-Y}^Y dy\,
f_q\left(\sqrt{\frac{\hat s}{s}}\,e^y\right)
f_{\bar q}\left(\sqrt{\frac{\hat s}{s}}\,e^{-y}\right),
\end{align}
with $Y=\half\log (s/\hat s)$, and where in the latter case,
a specific quark flavor $q$ is considered.
The quark convolution integrals enter in the cross section weighted
with different coupling constants and summed over quark flavors:
\begin{alignat}{2}
I_{BB}(\hat s)
&=2\sum_q I_{q\bar q}(\hat s), &\quad
I_{C_\gamma C_\gamma}(\hat s)&=2\sum_q Q_q^4 I_{q\bar q}(\hat s),\nonumber \\
I_{C_\gamma C_Z}(\hat s)&=2\sum_q Q_q^3 v_q I_{q\bar q}(\hat s), &\quad
I_{C_Z C_Z}(\hat s)
&=2\sum_q Q_q^2(v_q^2+a_q^2) I_{q\bar q}(\hat s), \nonumber \\
I_{D_\gamma D_\gamma}(\hat s)
&=2\sum_q Q_q^2 I_{q\bar q}(\hat s),  &\quad
I_{D_\gamma D_Z}(\hat s)
&=2\sum_q Q_q v_q I_{q\bar q}(\hat s),\nonumber \\
I_{D_Z D_Z}(\hat s)
&=2\sum_q (v_q^2+a_q^2) I_{q\bar q}(\hat s), &\quad
I_{C_\gamma D_Z}(\hat s)&=
I_{C_Z D_\gamma}(\hat s)=
2\sum_q Q_q^2 a_q I_{q\bar q}(\hat s), \nonumber \\
I_{C_Z D_Z}(\hat s)&=2\sum_q Q_q v_q a_q I_{q\bar q}(\hat s).
\end{alignat}

These integrals are labeled according to the sets of Feynman diagrams which
are associated with the couplings involved.  Integrals involving SM couplings
also enter in interference terms involving graviton exchange.  Note the factor
of two in the $q \bar q$ convolutions, which accounts for the fact that at a
$pp$ collider, either beam can provide the quark or the antiquark.  All
convolution integrals have been evaluated with CTEQ5 parton distribution
functions \cite{Lai:1999wy}.

The event shape distributions of the different contributions 
to the cross section can be expressed
in terms of $x_1$, $x_2$ and $x_3=1-x_1-x_2$.
It is convenient to express these quantities in terms of
the abbreviations:
\begin{align}
z_{a} &= 8x_3^4 - 12x_3^2 + 12x_3 -3, \nonumber\\
z_{b} &= 3x_3^2(1-2x_3)(2x_3^2 - 2x_3 + 1), \nonumber\\
z_{c} &= 2x_3^4(x_1+x_2)^2(4x_3^2 - 2x_3 +1), \nonumber\\
z_{d} &= 6(1-2x_3)(4x_3^2 - 10x_3 + 5), \nonumber\\
z_{e} &= 9x_3^2(1-2x_3)(2x_3^2 - 6x_3 + 3), \nonumber\\
z_{f} &= 8x_3^4 - 80x_3^3 + 180x_3^2 - 140x_3 + 35, \nonumber \\
z_{g} &= 2x_3^2 + 2x_3 - 1, \nonumber\\
z_{h} &= 2x_3^2 - 6x_3 + 3, \nonumber\\
z_{i} &= 8x_3^2 - 4x_3 + 3, \nonumber \\
z_{j} &= 8x_3^2 - 10x_3 + 3, \nonumber\\
z_{k} &= 12x_3^2 - 8x_3 + 3.
\end{align}

We first give the expression for initial-state radiation, expressed as
an integral over $\cos\theta$:
\begin{equation}
\label{Eq:XA}
X_A(x_1,x_2)=\int d(\cos\theta) \,
\frac{a_0(x_1,x_2) + a_2(x_1,x_2)\cos^2\theta
                 +a_4(x_1,x_2)\cos^4\theta}
{x_3^6(1-2x_3)(1-\cos^2\theta)},
\end{equation}
with
\begin{align}
a_0(x_1,x_2) &= - (x_1-x_2)^4 z_{a} - (x_1-x_2)^2 z_{b} + z_{c}, \nonumber \\
a_2(x_1,x_2) &= - (x_1-x_2)^4 z_{d} + (x_1-x_2)^2 z_{e} - x_3^2 z_{b},
\nonumber\\
a_4(x_1,x_2)
&= (x_1-x_2)^4 z_{f} - (x_1-x_2)^2 x_3^2 z_{d} - x_3^4 z_{a}.
\end{align}
For the initial-state SM background terms we have:
\begin{equation}
\label{Eq:XC}
X_{C}(x_1,x_2)=\int d(\cos\theta) \,
\frac{c_0(x_1,x_2) + c_2(x_1,x_2)\cos^2\theta}
     {x_3^4(1-\cos^2\theta)},
\end{equation}
with
\begin{align}
c_0(x_1,x_2) &=  (x_1-x_2)^2 z_{g} + x_3^2 z_{h}, \nonumber\\
c_2(x_1,x_2) &=  (x_1-x_2)^2 z_{h} + x_3^2 z_{g}.
\end{align}
We impose a cut on the photon emission angle $\theta$,
given by $|\eta|<2.5$, to render the integrals $X_A$ and $X_C$ finite.

For the remaining quantities, the integration over $\cos\theta$
is trivial, and one finds
\begin{align}
\label{Eq:XB}
X_B(x_1,x_2)
&=X_D(x_1,x_2)+\frac{8(x_1^2+x_2^2)}{1-2x_3}, \nonumber\\
X_D(x_1,x_2)&=\frac{2(x_1^2+x_2^2)}{(1-2x_1)(1-2x_2)}, \nonumber \\
X_{AD}(x_1,x_2)
&=(1-2x_3)\frac{(x_1-x_2)^2 z_{i}+ x_3^2 z_{j}}{x_3^4}, \nonumber \\
X_{BC}(x_1,x_2)
&=\frac{(x_1-x_2)^2 z_{k} + x_3^2(3-2x_3)}{x_3^4(1-2x_3)}, \nonumber \\
X_{CD}(x_1,x_2)&=\frac{x_1+x_2}{x_3^2}.
\end{align}
\section*{Appendix B: Definition of y-cuts}
\setcounter{equation}{0}
\renewcommand{\thesection}{B}

Here we shall define quantities where the $x_1,x_2$
dependence (which determines event shapes) in the cross sections is integrated
out.
When we carry out these integrations, we will impose $y$-cuts. In the case of
gluon--gluon fusion, we define
\begin{equation}
\label{Eq:y-cut}
S^{(G)}_{gg\to ee \gamma}=\iint\limits_{s_i>y\hat s} dx_1dx_2\,X_B(x_1,x_2).
\end{equation}
These $y$-cuts will remove IR soft and collinear events where the photon has
little energy, or its direction is close to that of an electron. The cut $y_3$,
which could be milder, will remove events where the two electrons are close.

In the case of quark--antiquark annihilation, the approach is the same, so we
will not write out the integrals here. However, for terms involving
initial-state radiation and related interference terms which depend on
$s_3=(1-2x_3)\hat s$, all factors involving $s_3$ must be part of the
integrand, since $x_1$ and $x_2$ are related to $x_3$.
\goodbreak


\clearpage
\end{document}